\documentclass[12pt,preprint]{aastex}

\shorttitle{Stellar Populations of Ellipticals I}
\shortauthors{Y.Yamada et al.}

\begin{document}

\title{Stellar Populations of Elliptical Galaxies in Virgo Cluster I. \\
The Data and Stellar Population Analysis \\}

\author{Y. Yamada\altaffilmark{1,2}}
\email{yamadays@cc.nao.ac.jp}
\author{N. Arimoto\altaffilmark{1}}
\author{A. Vazdekis\altaffilmark{3}}
\and
\author{R. F. Peletier\altaffilmark{4} \\}

\altaffiltext{1}{National Astronomical Observatory of Japan, 
2-21-1 Osawa, Mitaka, Tokyo 181-8588, Japan}
\altaffiltext{2}{School of Science, University of Tokyo, 
7-3-1 Hongo, Bunkyo, Tokyo 113-0033, Japan}
\altaffiltext{3}{Instituto de Astrof\'isica de Canarias, 
Via Lactea, E-38200 La Laguna, Tenerife, Spain}
\altaffiltext{4}{Kapteyn Astronomical Institute, University of Groningen, 
Postbus 800, 9700 AV Groningen, The Netherlands}

\begin{abstract}
We have determined spectroscopic ages of elliptical galaxies in the Virgo
cluster using spectra of very high signal-to-noise ratio (S/N~\AA
$^{-1}>$100). We observed 8 galaxies with
the Subaru Telescope and have combined this sample with 6 galaxies
previously observed with the WHT. To determine their ages we have used a new 
method based on the H$\gamma_\sigma$ age indicator, which is virtually independent of
the effects of metallicity. Apart from ages we have estimated abundances of
various elements. In this paper we present the observations, the data
reduction and the reliability of the H$\gamma_\sigma$ method. The results of
this investigation are presented in a companion paper (Yamada et al. 2006).
\end{abstract}


\section{Introduction}
An important issue in current-day astronomy is whether the stellar populations
in elliptical galaxies are uniformly old or a complex mixture of various
populations of different ages. One of the classical formation scenarios, the
monolithic collapse scenario (Eggen, Lynden-Bell \& Sandage 1962) suggests that
all elliptical galaxies consist of uniformly old stars : i.e., they formed at
high redshift ($z_f>$2) and evolved passively. The fundamental plane
(Djorgovski \& Davis 1987; Dressler et al. 1987; Bender, Burstein \& Faber
1993) and the Mg$_2-\sigma$ relation (Bender et al. 1993; Ziegler
\& Bender 1997; Colless et al. 1999) show indications towards this possibility. On the other hand,
the hierarchical clustering scenario predicts that elliptical galaxies
were formed as a result of merging of galaxies accompanied by intensive
starbursts (e.g. Toomre \& Toomre 1972; Kauffman \& Charlot 1998). Results 
based on spectroscopic line strength analysis by several authors (e.g. Gonz\'alez 1993; Trager et al. 2000b)
seem to support this scenario, since a significant scatter in ages is found. 
Recent results from the SLOAN survey (Kauffmann et al. 2003) seem to indicate that
large ellipticals are old and small ellipticals are younger. Although 
measuring accurate
ages and metallicities would lead to a breakthrough in our understanding of the
formation of elliptical galaxies, this has been considerably difficult due
to the age-metallicity degeneracy (Worthey 1994; Arimoto 1996). Photometric
and spectroscopic properties such as colors, absorption line indices, change
almost in the same way to differences in age and metallicity of the stellar 
populations, so that it is difficult to determine unambiguous ages.

In understanding the formation of elliptical galaxies in clusters, the
color-magnitude relation (CMR) plays a fundamental role. The CMR is an
important tool, since in clusters it is tight and as a result provides a 
relation that can be used to tie together ages, metallicities and masses of
galaxies. The scatter in the CMR is known to be small ($\sim$ 0.05 mag in the
relation between $V$ and $U-V$ (see Bower, Lucey \& Ellis 1992a,b; Terlevich et
al. 2001 for low redshift galaxies), and is also established in clusters at
high redshift (Stanford, Eisenhardt \& Dickinson 1998; Blakeslee et al. 2003),
supporting the monolithic collapse scenario. The origin of the CMR has been
explored extensively. A model based on the monolithic collapse scenario with a
galactic wind has been able to explain the CMR as a sequence of metallicity 
(Arimoto \& Yoshii 1987; Kodama \& Arimoto 1997).  This is not the only way to
explain the CMR: models based on hierarchical clustering (Kauffmann \& Charlot
1998; Ferreras, Charlot \& Silk 1999; Terlevich et al. 1999; Ferreras \& Silk
2000) producing a combination of age and metallicity variations could also explain
the CMR. This controversial situation is almost due to the
age-metallicity degeneracy. In short, the origin of the CMR is still under
debate. 

Spectroscopic properties, such as absorption line strengths, can partially
break the age-metallicity degeneracy. Worthey (1994) showed that the H$\beta$
index, defined in Burstein et al. (1984), is sensitive to the
luminosity-weighted age of galaxies and less to their metallicity. If a galaxy
contains ionized gas, however, its emission makes the H$\beta$ absorption line
weaker, and therefore an uncertain correction is necessary. To partially get
around this one can use H$\gamma$ and H$\delta$ indices, such as H$\gamma_\mathrm{A}$,
H$\delta_\mathrm{A}$, H$\gamma_\mathrm{F}$ and H$\delta_\mathrm{F}$ (defined in Jones \& Worthey 1995
and Worthey \& Ottaviani 1997), which are less affected by emission than
H$\beta$.  These aforementioned indices are some of the Lick/IDS indices 
(Burstein et al. 1984, Worthey et al. 1994), which are measuring line strengths
of several absorption lines at low resolution. Although maybe less than colors,
these indices are still affected by the age-metallicity degeneracy.

Partially to overcome the problem of low resolution Rose (1984) investigated 
spectra of elliptical galaxies using higher resolution indices ($\sim$2 \AA). 
His indices, such as H$\delta$/Fe I and
H$\gamma$/4325, are defined as ratios of neighboring absorption lines. Jones
\& Worthey (1995) defined a new H$_{\gamma}$ index, H$\gamma_{HR}$ that would
be more sensitive to age. That index, however, is strongly affected by
velocity dispersion ($\sigma$) broadening of galaxy spectra, since the bandpasses of
feature and pseudocontinua are very narrow, and therefore age determination using
this index are difficult and has uncertainty. Later, Vazdekis \& Arimoto (1999) introduced a
new age indicator, H$\gamma_\sigma$, which is very successful at
breaking the age-metallicity degeneracy and usable for a wide
range of $\sigma$. While the presence of non-solar abundance ratio in most
early-type galaxies (e.g., Gonz\'alez 1993; Kuntschner 2000; Poggianti et al.
2001, etc.) introduces systematic errors in the derived ages if
Lick indices are used, errors from H$\gamma_\sigma$ are less affected by 
non-solar abundance; i.e. errors mainly come from Poissonian noise. 

Vazdekis et al. (2001a) applied H$\gamma_\sigma$ to 6 elliptical galaxies in
the Virgo cluster and demonstrated that it is a useful age-determinator for
elliptical galaxies. The derived ages are uniformly old ($>$10Gyr), except for
one galaxy (NGC~4478) which has bluer color, and the total metallicities and
[Mg/Fe] ratios are increasing with luminosity, color and velocity
dispersion. They concluded that the CMR is driven mainly by the mean stellar
metallicity. Although this is an important result, it needs to be
strengthened, since the number of galaxies in Vazdekis et al. (2001a) is 
small, and the luminosity spans only 3.3 magnitudes.

In this paper we present new data of 8 elliptical galaxies in the Virgo cluster
observed with the Subaru Telescope together with the 6 galaxies of Vazdekis et
al. (2001a). For the combined sample of 14 galaxies we present the analysis of
the line strengths and determine their ages in various ways. We describe the
details of our sample selection and observations in Section 2. In Section 3,
we describe the line index measurements and stellar population analysis based on
the simple stellar population (SSP) models of Vazdekis (1999). In Section 4,
we present the measurement of ages and metallicities of our sample, and discuss 
the advantage of using H$\gamma_\sigma$ index and compare our results with
previous studies. In Section 5 the conclusions are given. In a future
paper (Yamada et al. 2006) we discuss age and metallicity trends along the
CMR and other scaling relations. 

\section{Observations and Sample Selection}
In this paper we present data taken with the Subaru and the William Herschel
Telescope. Vazdekis et al. (2001a) presented the first results of the WHT long
slit observations. We re-reduced the WHT data to check the dependence 
of data reduction technique on the index measurements and added 8 new galaxies 
observed with the Subaru Telescope. 

\subsection{Subaru Observation} The Subaru observations were done on April
17-18, 2002 and June 26, 2003. The spectroscopic observations were performed
with the long slit mode of the FOCAS spectrograph with the MIT CCD and the R300
grism in the 2nd order (Kashikawa et al. 2002). This instrumental setup provides a
dispersion of 0.67 \AA pixel$^{-1}$ and a spectral range of
$\lambda\lambda\simeq$ 3800\AA--5800\AA. We used a slit width of 0$\arcsec$.4,
giving a resolution of 2.0 \AA\ ($\sigma\sim60$ km s$^{-1}$) on 
April 17, 2002 and a slit width of 0$\arcsec$.6, corresponding to a resolution
of 3.1 \AA\ ($\sigma\sim95$ km s$^{-1}$) on April 18, 2002 and
June 26, 2003. The spatial scale was 0$\arcsec$.21 pix$^{-1}$ after binning 2
pixels and the seeing was 0$\arcsec$.6--1$\arcsec$.0 for all nights. Multiple
exposures of 20 minutes length were acquired for each galaxy, as well as ThAr
arc lamp exposures for wavelength calibration at every galaxy pointing. We
obtained a number of dome and twilight flat fields for flatfielding. We also
observed one spectrophotomeric standard star per run, Feige~34 in 2002 April
and BD+33~2642 in 2003 June. 

\subsection{WHT Observation} The WHT observations were done in April 21-22,
1999. We used the ISIS double-beam spectrograph with the EEV12 CCD and R600B
grating in the blue channel. The dispersion and spectral resolution is 0.44
\AA pixel$^{-1}$ and 2.4 \AA\ ($\sigma\sim65$ km s$^{-1}$) provided by a slit
width of 1$\arcsec$.6 and a spectral range of $\lambda\lambda\simeq$
4000--5500\AA. The seeing was 2$\arcsec$--3$\arcsec$ during the first night and
$\sim$ 1$\arcsec$.5 during the second. Multiple exporsures of 35 minutes
length were obtained for each galaxy. We made use of tungsten 
flat fields obtained at each galaxy pointing and twilight flat fields as well
as CuArNe arc lamp frames for $\lambda$ calibration for each exposure. For
correction to a relative flux scale the spectrophotometric standard star 
G191B29 was used. 

\subsection{Data Reduction} Data reduction was done with the standard IRAF
packages in the standard way; i.e. overscan correction, bias subtraction, flat
fielding, $\lambda$ calibration, sky subtraction and flux calibration. We used
dome flat fields (for the Subaru data) and tungsten lamp flat fields 
(for the WHT data)
for correction for pixel-to-pixel variations of the sensitivity. 
The spectra of
WHT have wiggles coming from problems with the dichroic. We successfully
removed these patterns by using tungsten lamp flat fields obtained for each
galaxy pointing. We achieved $\Delta\lambda$ (error in $\lambda$) smaller than
0.07\AA\ for the $\lambda$ calibration, which is very important for the
determination of H$\gamma_\sigma$, since even a small shift in $\lambda$ can
affect the H$\gamma_\sigma$ measurement significantly (Vazdekis \& Arimoto
1999). Cosmic rays were removed using the "cleanest" task of the REDUCEME
package (Cardiel et al. 1998). Exposures whose spectra around H$\gamma$
feature were hit by cosmic ray were excluded. Flux calibration
was done with the observed flux standard stars G191B29 (for the WHT data),
Feige~34 and BD+33~2642 (for the Subaru data). 

In this study, since we are interested in the central part of galaxies, we have
summed up the spectra within $r_e$/10. Most of elliptical galaxies are known to
have color and line strength gradients (e.g., Peletier et al. 1990, Gonz\'alez 
1993; Kobayashi \& Arimoto 1999). Here we used effective radii from the RC3
catalogue (de Vaucouleurs et al. 1991).  Before summing the central galaxy
spectra, we carefully aligned them with the rows of the array, and each row in
spatial direction was corrected for the velocity shift due to the rotation by
cross-correlating with the Vazdekis (1999) models. Variations of
$\sigma$ within $r_e/10$ were small ($<$ 25 km s$^{-1}$) for all galaxies, and
therefore no correction for this effect was applied. 

To check for systematic effects due to telescope characteristics and data
reduction, we compared the line indices for NGC~5831, a galaxy observed during
both Subaru and WHT runs. We found excellent agreement between these indices measurements
(see the open triangle (WHT) and open diamond (Subaru) in the middle panel of
Fig.~\ref{Hbalmer-MgFe-V}). Moreover, WHT data reduced independently by two
authors (Y.Yamada. and A.Vazdekis) show also good agreement. 

\subsection{Sample Selection} Table~\ref{ObsGal-V} lists the basic parameters
of the observed galaxies, such as radial velocity (V$_r$), effective radius
($r_e$), velocity dispersion ($\sigma$), visual magnitude ($m_V$), $U-V$ and
$B-V$ colors, telescope, exposure time, slit width, and signal-to-noise
ratio per Angstrom around 4340\AA. 

\begin{figure*}[!h]
\epsscale{0.8}
\plotone{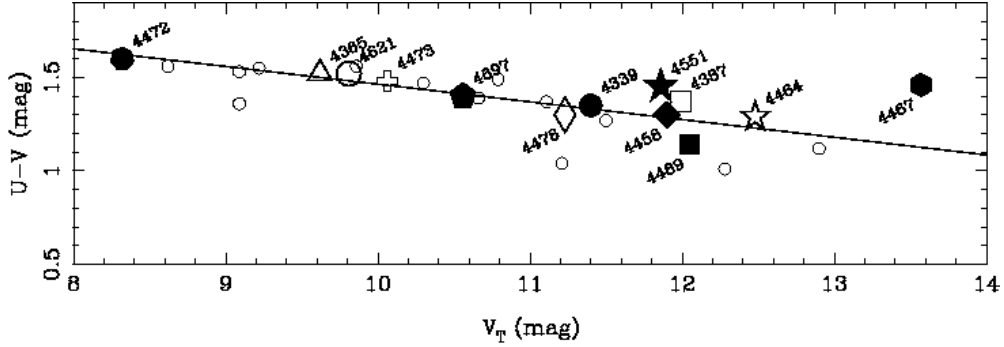}
\caption{
\footnotesize
Color-magnitude relation of Virgo cluster elliptical galaxies (Bower et al. 1992a,b). 
Our sample is indicated by the large symbols with NGC numbers. 
NGC~4239, whose symbol is a filled triangle, has no $U-V$ data. 
\normalsize
\label{CMR-BLE92}}
\end{figure*}

Our sample was primarily based on the Virgo galaxies observed by Michard (1982)
and Bower et al. (1992a,b), which show the CMRs with surprisingly
small scatter. We included 3 elliptical galaxies from Jones (1998) in order to
enlarge our luminosity coverage toward the fainter end. In Fig.~\ref{CMR-BLE92}
we reproduce the CMR of our sample galaxies. We selected the galaxies
according to the following criteria: 1) Since we were only interested in
elliptical galaxies, we excluded galaxies classified as S0 in the references
mentioned above and the RC3 catalogue (de Vaucouleurs et al. 1991). 
2) We excluded galaxies having
strong emission lines, such as NGC~4374 and NGC~4486. 3) We selected galaxies
evenly for each luminosity bin $m_B<11.5$, $11.5<m_B<12.5$, $12.5<m_B<13.5$,
and $m_B>13.5$, to cover a wide range of the luminosity. The luminosity spans
5.8 magnitudes, which is larger than the data of Vazdekis et al. (2001a), which
spans only 3.3 magnitudes.


\begin{table}[!h]
\begin{center}
\caption{Properties and Observational Parameters of our Sample Ellipticals \label{ObsGal-V}}
\scriptsize
\begin{tabular}{lrrrrrrcrrr}
\\
\hline\hline
NGC & V$_r$       & $r_e$  & $\sigma$    & $m_V$   & $U-V$ & $B-V$ & Obs. & Exp.Time & Slit  & S/N        \\
    & km s$^{-1}$ & arcsec & km s$^{-1}$ & mag     & mag   & mag   &      & min.     &       & \AA $^{-1}$\\
    & (1)         & (2)    & (3)         & (4)     & (5)   & (6)   & (7)  & (8)      & (9)   & (10)       \\
\hline
NGC4239 &   921  &    15  &    63  &  12.81 &      - &  0.89  &  S1    &   100  &  0$\arcsec$.4  &   115 \\
NGC4339 &  1281  &    32  &   114  &  11.73 & 1.35   &  0.92  &  S2    &    50  &  0$\arcsec$.6  &   150 \\
NGC4365 &  1227  &    80  &   261  &   9.67 & 1.52   &  0.99  &  W     &    70  &  1$\arcsec$.6  &   250 \\
NGC4387 &   561  &    16  &   112  &  12.29 & 1.37   &  0.91  &  W     &   105  &  1$\arcsec$.6  &   130 \\
NGC4458 &   668  &    28  &   106  &  12.33 & 1.30   &  0.88  &  S1    &    70  &  0$\arcsec$.4  &   145 \\
NGC4464 &  1255  &     3  &   121  &  12.93 & 1.29   &  0.86  &  W     &   105  &  1$\arcsec$.6  &   160 \\
NGC4467 &  1426  &     6  &    83  &  14.33 & 1.46$^a$ & 0.86 &  S3    &   150  &  0$\arcsec$.6  &   115 \\
NGC4472 &   983  &   107  &   303  &   8.55 & 1.60   &  0.98  &  S1    &    20  &  0$\arcsec$.4  &   170 \\
NGC4473 &  2237  &    42  &   193  &  10.33 & 1.48   &  1.02  &  W     &    70  &  1$\arcsec$.6  &   280 \\
NGC4478 &  1382  &    13  &   143  &  11.46 & 1.32$^b$ & 0.93 &  W     &   105  &  1$\arcsec$.6  &   135 \\
NGC4489 &   957  &    28  &    62  &  12.38 & 1.21$^a$ & 0.89 &  S2    &    40  &  0$\arcsec$.6  &   120 \\
NGC4551 &  1189  &    18  &   113  &  12.14 & 1.45   &  0.92  &  S2    &    80  &  0$\arcsec$.6  &   160 \\
NGC4621 &   430  &    44  &   230  &   9.81 & 1.52   &  0.96  &  W     &   140  &  1$\arcsec$.6  &   490 \\
NGC4697 &  1236  &    72  &   181  &  10.56 & 1.40   &  0.89  &  S1    &    30  &  0$\arcsec$.4  &   255 \\
\hline
\end{tabular}
\end{center}
\footnotesize
(1): RC3, de Vaucouleurs et al. (1991); (2)(4)(6): RC3 ; (3): McElroy (1995); 
(5): Bower et al. (1992a), $^{a}$RC3 catalogue, $^{b}$Michard (1982), 
(7): S1, S2, S3 and W means observation with Subaru (1st, 2nd, 3rd night) and
WHT, respectively; (8)(9): Exposure time and slit width used; (10):
S/N~\AA $^{-1}$ at 4340\AA, the wavelength of H$\gamma$ 
\normalsize
\end{table}


\subsection{Line Indices}
We used the H$\gamma_\sigma$ index (Vazdekis \& Arimoto 1999; Vazdekis et al.
2001b) and the indices of the Lick system (Worthey 1994; Jones \& Worthey
1995; Worthey \& Ottaviani 1997). For measuring the indices we used the FORTRAN 
program called ``LECTOR" \footnote{See
http:///www.iac.es/galeria/vazdekis/models.html}, which is based on the
equations (41)--(44) and simulations in Cardiel et al. (1998). 


\begin{table}[!t]
\begin{center}
\caption{H$\gamma_\sigma$ Indices \label{DefHgamma}}
\footnotesize
\begin{tabular}{ccccccc}
\hline\hline
Name     & Blue Continuum & Index Bandpass & Red Continuum
         & Units          & $\sigma_{measured}$ $^b$ & Required \\
         & $\lambda_{CB1}$--$\lambda_{CB2}$ 
         & $\lambda_{I1}$ --$\lambda_{I2}$ 
         & $\lambda_{CR1}$--$\lambda_{CR2}$ 
         &                & km s$^{-1}$    & S/N~\AA $^{-1}$ $^a$ \\
\hline
H$\gamma_{\sigma\le130}$ & 4329.000--4340.468 & 4333.250--4363.000 & 4352.500--4368.250 
                   & \AA &  60--130 & 200 \\
H$\gamma_{125}$    & 4330.000--4340.468 & 4333.000--4352.737 & 4359.250--4368.750 
                   & \AA & 100--175 & 200 \\
H$\gamma_{200}$    & 4331.000--4340.750 & 4332.000--4352.250 & 4359.250--4368.750 
                   & \AA & 150--225 & 300 \\
H$\gamma_{275}$    & 4331.500--4341.000 & 4331.500--4351.875 & 4359.250--4368.750 
                   & \AA & 225--300 & 400 \\
\hline
\end{tabular}
\end{center}
\footnotesize
{}$^a$ Required S/N~\AA $^{-1}$ to distinguish between 12 Gyrs and 17.5 Gyrs 
stellar population (Vazdekis \& Arimoto 1999). \\
{}$^b$ Each H$\gamma_\sigma$ index can be applied in this $\sigma_{\mathrm{measured}}$ 
($=(\sigma^2_{\mathrm{galaxy}}+\sigma^2_{instr}+\sigma^2_{\mathrm{extra}})^\frac{1}{2}$) range.
\normalsize
\end{table}


We list the wavelength definition of 4 H$\gamma_\sigma$ indices in
Table~\ref{DefHgamma}. Since the required S/N~\AA $^{-1}$ to distinguish
between models of 12 Gyrs and 17.5 Gyrs is very high (S/N~\AA $^{-1}$ = 200 for
H$\gamma_{\sigma\le130}$ and S/N~\AA $^{-1}$ = 400 for H$\gamma_{275}$), we
integrated for a considerable time and consequently reached sound-to-noise ratios 
\AA $^{-1}$ ranging from 115 to 490 for the selected apertures, i.e., $r_e/10$. 
Although the S/N~\AA $^{-1}$ of our data is
slightly lower than listed in Table~\ref{DefHgamma} in most galaxies, we could
achieve a high precision in the age determination for the following reasons: 
1) According to Cardiel et al. (2003), the real random errors of
H$\gamma_\sigma$ are smaller than those computed by the LECTOR
program since the wavelength definitions of H$\gamma_\sigma$ are not usual, 
i.e., the pseudo-continua and the feature bandpasswd overlap. We used the real error by
using their new estimate of the random errors for H$\gamma_\sigma$. 2) The
H$\gamma_\sigma$ index decreases rapidly for young ages, just as other
Balmer lines do. Although the S/N~\AA $^{-1}$ for NGC4489 is low ($\sim$ 115),
the error in age is only 0.6 Gyrs.

The spectral range of our data goes from 3800\AA\ to 5800\AA\ to cover many of
the Lick indices: H$\delta_\mathrm{A}$ and H$\gamma_\mathrm{A}$ (Jones \& Worthey 1995),
H$\delta_\mathrm{F}$ and H$\gamma_\mathrm{F}$ (Worthey \& Ottaviani 1997), H$\beta$, CN$_1$,
CN$_2$, Ca4227, G4300, Fe4383, Ca4455, Fe4531, C4668, Fe5015, Mg$_1$, Mg$_2$,
Mg$b$, Fe5270, Fe5335, Fe5406, Fe5709 and NaD (Worthey et al. 1994). Worthey
(1994), Jones \& Worthey (1995) and Worthey \& Ottaviani (1997) have investigated the age and metallicity
sensitivity of these indices. They concluded that Balmer line indices
are sensitive to age more than metallicity. Besides H$\gamma_\sigma$ we used
H$\beta$ and H$\delta_\mathrm{F}$ as age indicators for reference. Concerning the
metallicity, we used [MgFe] (defined in Gonz\'alez 1993) to estimate the mean
metallicity [M/H], 
\[
[\textrm{MgFe}]=\sqrt{\textrm{Mg}b\cdot(\textrm{Fe}5270+\textrm{Fe}5335) / 2}\;. 
\] 
The feasibility of using
[MgFe] for the total metallicity is shown in Vazdekis et al. (2001a), Thomas,
Maraston \& Bender (2003), and Bruzual \& Charlot (2003). 
Apart from this, we used several indices to determine abundances of
several elements and molecules. For Fe we use the
Fe3 index defined in Kuntschner (2000),
\[
\textrm{Fe}3=(\textrm{Fe}4383+\textrm{Fe}5270+\textrm{Fe}5335)/3\;. 
\]
to give an Fe abundance estimate [Z$_{\mathrm{Fe}}$/H]. 
Mg$b$, Ca4227, CN$_2$ and G4300 are used to determine [Z$_{\mathrm{Mg}}$/H], 
[Z$_{\mathrm{Ca}}$/H], [Z$_{\mathrm{CN}}$/H] and [Z$_{\mathrm{CH}}$/H], respectively. 
Tripicco \& Bell (1995) and Trager et al. (1998) confirmed the
dominant element or molecule in each line index. But each index is not pure, 
because the indices are still somewhat contaminated by other elements or molecules. 
For the G-band, Worthey (1994) and Tripicco \& Bell (1995) show us that 
the contributers to the G-band are CH, Fe and Ti. Yet, since the main contributer 
is still CH (C), we write [Z$_{\mathrm{CH}}$/H] for convenience. 

Note that, although the bandpasses are mostly those of the Lick system, we 
{\sl did not use the Lick/IDS system itself}, instead we used flux-calibrated
spectra and no offsets were applied to reproduce the stars of the Lick/IDS system. 
Measured line strength of the indices are listed in Table~\ref{Index1-V} \&
\ref{Index2-V}. 


\begin{deluxetable}{cccccccccccccccc} 
\rotate
\tabletypesize{\scriptsize}
\tablecolumns{16} 
\tablewidth{0pc} 
\tablecaption{Line Indices (Blue)\label{Index1-V}} 
\startdata 
\hline\hline
Galaxy  & $\sigma_{gal}$$^a$                      & $\sigma_{meas}$$^b$
        & H$\gamma_{\sigma<130}$$^c$        & H$\gamma_{125}$$^d$ 
        & H$\gamma_{200}$$^d$               & H$\gamma_{275}$$^d$ 
        & H$\delta_\mathrm{A}$     & H$\gamma_\mathrm{A}$     & H$\delta_\mathrm{F}$     & H$\gamma_\mathrm{F}$ 
        & CN$_1$          & CN$_2$          & Ca4227          & G4300 
        & Fe4383 \\
Error   & \multicolumn{2}{c}{km s$^{-1}$} 
        & (\AA)           & (\AA)           & (\AA)           & (\AA) 
        & (\AA)           & (\AA)           & (\AA)           & (\AA) 
        & (mag)           & (mag)           & (\AA)           & (\AA) 
        & (\AA) \\
\\
\hline
NGC4239&  82& 150&  1.213 &  0.996 &    -   &    -   & -0.959 & -4.044 &  0.909 & -0.517 &  0.008 &  0.046 &  1.632 &  5.130 &  4.829 \\
 $\pm$ &    &    &  0.067 &  0.040 &    -   &    -   &  0.103 &  0.082 &  0.069 &  0.051 &  0.003 &  0.003 &  0.044 &  0.074 &  0.100 \\
NGC4339& 142& 150&     -  &  0.836 &    -   &    -   & -2.602 & -6.154 &  0.323 & -1.851 &  0.091 &  0.138 &  1.717 &  5.476 &  5.848 \\
 $\pm$ &    &    &     -  &  0.030 &    -   &    -   &  0.087 &  0.064 &  0.057 &  0.041 &  0.002 &  0.002 &  0.034 &  0.056 &  0.072 \\
NGC4365& 245& 300&     -  &    -   &    -   &  0.252 & -2.820 & -7.875 & -0.292 & -2.681 &  0.121 &  0.163 &  1.740 &  5.815 &  5.516 \\
 $\pm$ &    &    &     -  &    -   &    -   &  0.017 &  0.039 &  0.042 &  0.026 &  0.026 &  0.001 &  0.001 &  0.019 &  0.033 &  0.049 \\
NGC4387& 105& 150&  0.961 &  0.827 &    -   &    -   & -2.666 & -6.360 &  0.253 & -1.909 &  0.048 &  0.093 &  1.763 &  6.067 &  5.949 \\
 $\pm$ &    &    &  0.061 &  0.036 &    -   &    -   &  0.074 &  0.078 &  0.048 &  0.049 &  0.002 &  0.002 &  0.035 &  0.062 &  0.092 \\
NGC4458& 104& 150&  0.919 &  0.823 &    -   &    -   & -2.109 & -5.352 &  0.472 & -1.441 &  0.074 &  0.121 &  1.519 &  5.573 &  4.830 \\
 $\pm$ &    &    &  0.051 &  0.030 &    -   &    -   &  0.083 &  0.063 &  0.055 &  0.040 &  0.002 &  0.002 &  0.034 &  0.056 &  0.075 \\
NGC4464& 135& 150&     -  &  0.778 &    -   &    -   & -2.781 & -6.240 &  0.107 & -1.828 &  0.106 &  0.156 &  1.285 &  5.886 &  5.422 \\
 $\pm$ &    &    &     -  &  0.028 &    -   &    -   &  0.059 &  0.061 &  0.038 &  0.038 &  0.002 &  0.002 &  0.028 &  0.049 &  0.074 \\
NGC4467&  75& 150&  0.999 &  0.830 &    -   &    -   & -1.792 & -5.257 &  0.425 & -1.429 &  0.110 &  0.151 &  1.586 &  5.389 &  5.273 \\
 $\pm$ &    &    &  0.055 &  0.033 &    -   &    -   &  0.082 &  0.069 &  0.054 &  0.043 &  0.002 &  0.002 &  0.036 &  0.059 &  0.082 \\
NGC4472& 306& 300&     -  &    -   &    -   &  0.303 & -2.937 & -6.762 & -0.167 & -1.959 &  0.131 &  0.177 &  1.578 &  5.341 &  5.330 \\
 $\pm$ &    &    &     -  &    -   &    -   &  0.026 &  0.092 &  0.065 &  0.061 &  0.041 &  0.002 &  0.002 &  0.035 &  0.057 &  0.073 \\
NGC4473& 180& 190&     -  &    -   &  0.498 &    -   & -3.529 & -6.993 & -0.086 & -2.303 &  0.138 &  0.190 &  1.586 &  5.908 &  5.945 \\
 $\pm$ &    &    &     -  &    -   &  0.015 &    -   &  0.035 &  0.036 &  0.022 &  0.022 &  0.001 &  0.001 &  0.016 &  0.028 &  0.042 \\
NGC4478& 132& 150&     -  &  0.889 &    -   &    -   & -2.078 & -5.786 &  0.512 & -1.559 &  0.077 &  0.123 &  1.668 &  5.564 &  5.874 \\
 $\pm$ &    &    &     -  &  0.034 &    -   &    -   &  0.071 &  0.072 &  0.047 &  0.045 &  0.002 &  0.002 &  0.033 &  0.060 &  0.086 \\
NGC4489&  73& 150&  1.275 &  1.093 &    -   &    -   & -1.384 & -4.264 &  0.800 & -0.631 &  0.024 &  0.069 &  1.695 &  5.294 &  5.523 \\
 $\pm$ &    &    &  0.072 &  0.043 &    -   &    -   &  0.124 &  0.089 &  0.083 &  0.056 &  0.003 &  0.003 &  0.050 &  0.082 &  0.105 \\
NGC4551& 105& 150&     -  &  0.928 &    -   &    -   & -2.871 & -6.162 &  0.271 & -1.729 &  0.084 &  0.129 &  1.901 &  5.645 &  6.176 \\
 $\pm$ &    &    &     -  &  0.027 &    -   &    -   &  0.075 &  0.058 &  0.049 &  0.036 &  0.002 &  0.002 &  0.030 &  0.050 &  0.065 \\
NGC4621& 230& 300&     -  &    -   &    -   &  0.299 & -3.100 & -7.021 & -0.064 & -2.202 &  0.154 &  0.202 &  1.438 &  5.649 &  5.499 \\
 $\pm$ &    &    &     -  &    -   &    -   &  0.008 &  0.020 &  0.020 &  0.013 &  0.013 &  0.001 &  0.001 &  0.009 &  0.016 &  0.024 \\
NGC4697& 168& 190&     -  &    -   &  0.493 &    -   & -3.028 & -6.311 &  0.029 & -1.943 &  0.116 &  0.159 &  1.701 &  5.598 &  5.672 \\
 $\pm$ &    &    &     -  &    -   &  0.016 &    -   &  0.048 &  0.036 &  0.031 &  0.023 &  0.001 &  0.001 &  0.019 &  0.031 &  0.041 \\
\hline
\enddata 
\tablenotetext{a} {\scriptsize
 Velocity dispersion measured from spectra. }
\tablenotetext{b} {\scriptsize
 Adopted total velocity dispersion for the measurements. }
\tablenotetext{c} {\scriptsize
 Measurements for $\sigma_{\mathrm{measured}}=130$ km s$^{-1}$ (NGC~4239, NGC~4387,
NGC~4458, NGC~4467 and NGC~4489).}
\tablenotetext{d} {\scriptsize
 For H$\gamma_\sigma$, we show the measurement only for the appropriate
 velocity dispersion $\sigma_{\mathrm{measured}}$ (150, 190 or 300 km s$^{-1}$).}
\end{deluxetable} 


\begin{deluxetable}{ccccccccccccccc} 
\rotate
\tabletypesize{\scriptsize}
\tablecolumns{15} 
\tablewidth{0pc} 
\tablecaption{Line Indices (Red) \label{Index2-V}} 
\startdata 
\hline\hline
Galaxy  & Ca4455          & Fe4531          & C4668 
        & H$\beta$        & Fe5015          & Mg$_1$          & Mg$_2$ 
        & Mg$b$           & Fe5270          & Fe5335          & Fe5406 
        & Fe5709          & Fe5782          & NaD \\
Error   & (\AA)           & (\AA)           & (\AA) 
        & (\AA)           & (\AA)           & (mag)           & (mag) 
        & (\AA)           & (\AA)           & (\AA)           & (\AA) 
        & (\AA)           & (\AA)           & (\AA) \\
\\
\hline
NGC4239&  1.612 &  3.456 &  4.839 &  2.204 &  5.497 &  0.031 &  0.150 &  3.135 &  2.856 &  2.641 &  1.746 &  1.029 &  0.819 &  1.793 \\
 $\pm$ &  0.050 &  0.073 &  0.109 &  0.043 &  0.096 &  0.001 &  0.001 &  0.048 &  0.054 &  0.063 &  0.047 &  0.042 &  0.042 &  0.061 \\
NGC4339&  1.818 &  6.097 &  0.098 &  0.251 &  4.495 &  3.200 &  3.065 &  2.090 &  1.064 &  1.015 &  4.229 &  1.785 &  3.584 &  7.859 \\
 $\pm$ &  0.036 &  0.052 &  0.075 &  0.031 &  0.067 &  0.001 &  0.001 &  0.033 &  0.037 &  0.042 &  0.032 &  0.028 &  0.028 &  0.040 \\
NGC4365&  1.544 &  4.090 &  8.207 &  1.467 &  5.670 &  0.137 &  0.301 &  4.810 &  2.886 &  2.552 &  1.754 &    -   &    -   &    -   \\
 $\pm$ &  0.026 &  0.039 &  0.060 &  0.025 &  0.057 &  0.001 &  0.001 &  0.027 &  0.031 &  0.035 &  0.026 &    -   &    -   &    -   \\
NGC4387&  1.703 &  3.632 &  5.927 &  1.662 &  5.897 &  0.096 &  0.234 &  4.200 &  3.141 &  2.964 &  2.004 &    -   &    -   &    -   \\
 $\pm$ &  0.049 &  0.074 &  0.116 &  0.048 &  0.105 &  0.001 &  0.001 &  0.051 &  0.057 &  0.065 &  0.049 &    -   &    -   &    -   \\
NGC4458&  1.678 &  5.080 &  0.068 &  0.206 &  4.241 &  2.799 &  2.607 &  1.671 &  0.966 &  0.897 &  2.686 &  1.513 &  3.392 &  5.040 \\
 $\pm$ &  0.037 &  0.054 &  0.080 &  0.032 &  0.070 &  0.001 &  0.001 &  0.034 &  0.039 &  0.045 &  0.034 &  0.030 &  0.029 &  0.042 \\
NGC4464&  1.609 &  3.458 &  5.103 &  1.599 &  5.332 &  0.104 &  0.248 &  4.425 &  2.829 &  2.506 &  1.745 &    -   &    -   &    -   \\
 $\pm$ &  0.039 &  0.059 &  0.093 &  0.038 &  0.084 &  0.001 &  0.001 &  0.040 &  0.046 &  0.053 &  0.039 &    -   &    -   &    -   \\
NGC4467&  1.733 &  5.620 &  0.099 &  0.243 &  4.234 &  3.121 &  2.897 &  2.022 &  1.026 &  0.918 &  3.233 &  1.712 &  3.482 &  5.909 \\
 $\pm$ &  0.042 &  0.062 &  0.092 &  0.037 &  0.082 &  0.001 &  0.001 &  0.040 &  0.045 &  0.052 &  0.039 &  0.034 &  0.034 &  0.048 \\
NGC4472&  1.580 &  5.246 &  0.124 &  0.289 &  4.545 &  2.749 &  2.375 &  1.607 &  0.827 &  0.812 &  4.854 &  1.317 &  3.474 &  8.169 \\
 $\pm$ &  0.036 &  0.052 &  0.074 &  0.030 &  0.067 &  0.001 &  0.001 &  0.034 &  0.037 &  0.042 &  0.032 &  0.029 &  0.028 &  0.040 \\
NGC4473&  1.727 &  3.650 &  8.711 &  1.632 &  5.906 &  0.136 &  0.294 &  4.884 &  3.182 &  2.886 &  1.986 &    -   &    -   &    -   \\
 $\pm$ &  0.022 &  0.034 &  0.052 &  0.022 &  0.049 &  0.001 &  0.001 &  0.024 &  0.026 &  0.030 &  0.023 &    -   &    -   &    -   \\
NGC4478&  1.856 &  3.733 &  6.568 &  1.919 &  6.220 &  0.104 &  0.253 &  4.435 &  3.244 &  3.117 &  2.075 &    -   &    -   &    -   \\
 $\pm$ &  0.045 &  0.068 &  0.106 &  0.043 &  0.095 &  0.001 &  0.001 &  0.046 &  0.051 &  0.058 &  0.043 &    -   &    -   &    -   \\
NGC4489&  2.413 &  6.171 &  0.046 &  0.173 &  3.456 &  3.305 &  3.068 &  1.973 &  1.141 &  1.006 &  2.656 &  1.678 &  3.626 &  6.345 \\
 $\pm$ &  0.051 &  0.074 &  0.108 &  0.042 &  0.093 &  0.001 &  0.001 &  0.047 &  0.052 &  0.060 &  0.046 &  0.041 &  0.040 &  0.059 \\
NGC4551&  1.896 &  6.203 &  0.091 &  0.245 &  4.527 &  3.371 &  3.260 &  2.177 &  1.156 &  1.055 &  3.468 &  1.809 &  3.719 &  7.538 \\
 $\pm$ &  0.033 &  0.048 &  0.071 &  0.029 &  0.063 &  0.001 &  0.001 &  0.031 &  0.035 &  0.040 &  0.030 &  0.026 &  0.026 &  0.038 \\
NGC4621&  1.447 &  3.593 &  8.958 &  1.480 &  5.560 &  0.165 &  0.325 &  4.909 &  2.943 &  2.645 &  1.783 &    -   &    -   &    -   \\
 $\pm$ &  0.013 &  0.019 &  0.029 &  0.012 &  0.028 &  0.000 &  0.000 &  0.013 &  0.015 &  0.017 &  0.013 &    -   &    -   &    -   \\
NGC4697&  1.603 &  5.869 &  0.108 &  0.261 &  4.591 &  3.224 &  3.003 &  2.037 &  1.019 &  0.999 &  4.220 &  1.643 &  3.709 &  8.194 \\
 $\pm$ &  0.020 &  0.030 &  0.043 &  0.018 &  0.039 &  0.000 &  0.000 &  0.019 &  0.021 &  0.025 &  0.019 &  0.016 &  0.016 &  0.023 \\
\enddata 
\end{deluxetable} 


\normalsize

\subsection{Stellar Population Models}
To derive the ages and metallicities of the galaxies, we used the revised 
evolutionary stellar population synthesis model of Vazdekis (1999), which has
been made using the empirical stellar spectral library of Jones (1998) and
updated with the new Padova isochrones (Girardi et al. 2000). 
The model predicts spectral energy distributions (SEDs) in the optical
wavelength range, $\lambda\lambda$3976\AA--4456\AA\ and
$\lambda\lambda$4764\AA--5460\AA, for simple stellar populations (SSPs) at a
resolution of 1.8\AA~(FWHM). Previous models (e.g., Worthey 1994; Vazdekis et
al. 1996) mostly used the Lick polynomial fitting functions for absorption
line indices (Worthey et al. 1994; Worthey \& Ottaviani 1997) which are based
on low resolution spectra (FWHM $\sim$ 9\AA), thus limiting the application
of the models to strong features. The new model provides flux-calibrated,
high-resolution spectra, therefore allowing an analysis of galaxy spectra at
the natural resolution given by their internal velocity broadening and
instrumental resolution. Thus no uncertain correction for
velocity broadening has to be applied as is the case when using the
Lick/IDS system. 


\begin{figure*}[!p]
\epsscale{1.0}
\plotone{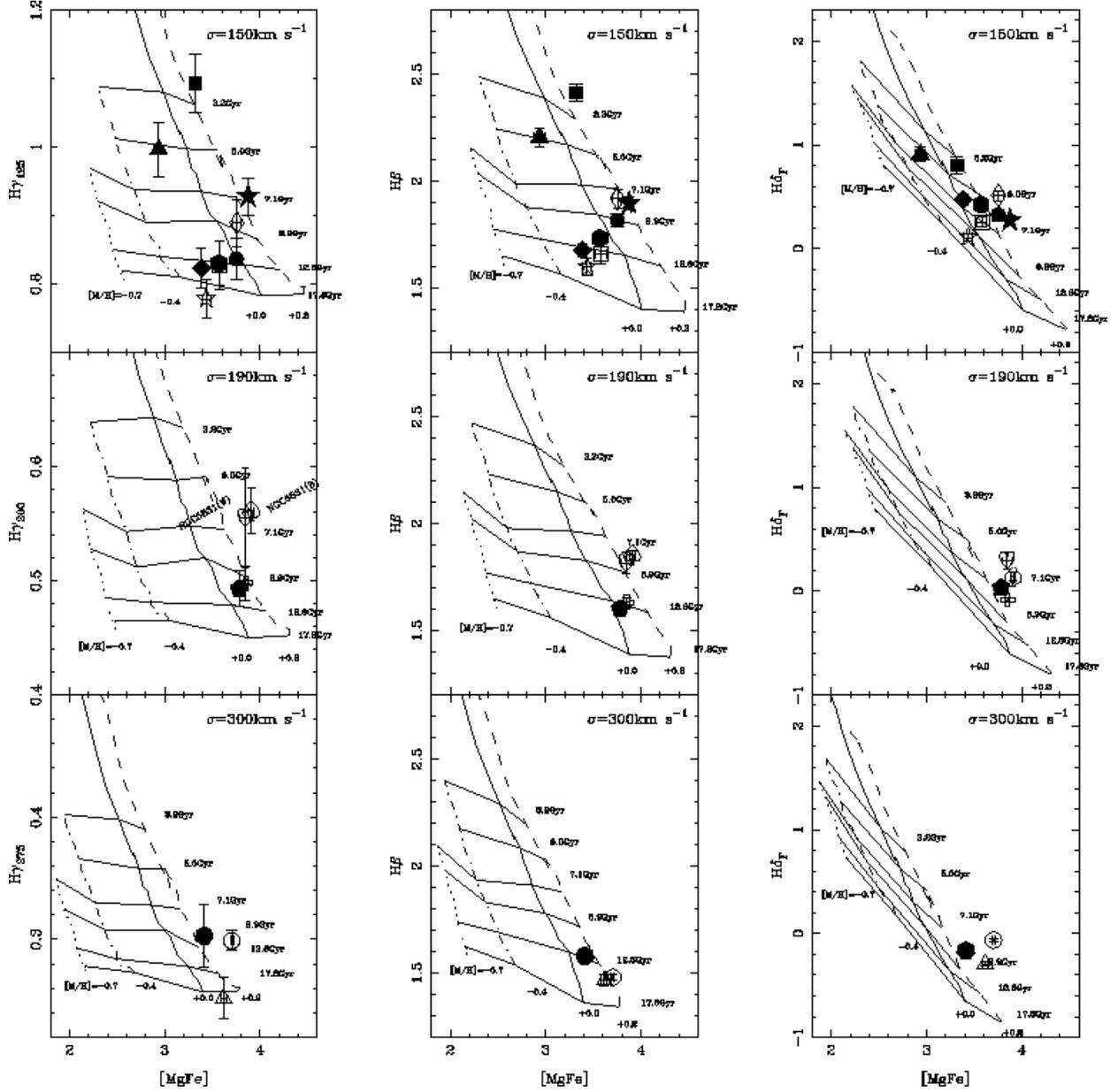}
\caption{
\footnotesize
H$\gamma_\sigma$, H$\beta$, H$\delta_\mathrm{F}$ vs. [MgFe] diagrams. 
Top, middle and bottom panels are for galaxies with $\sigma_{\mathrm{measured}}=150$ km
s$^{-1}$, 190 km s$^{-1}$ and 300 km s$^{-1}$, respectively. Model grids with
various ages and metallicities are overplotted. Symbols are the same as in 
Fig.~\ref{CMR-BLE92}.
\normalsize
\label{Hbalmer-MgFe-V}}
\end{figure*}

\begin{figure*}[!p]
\epsscale{1.0}
\plotone{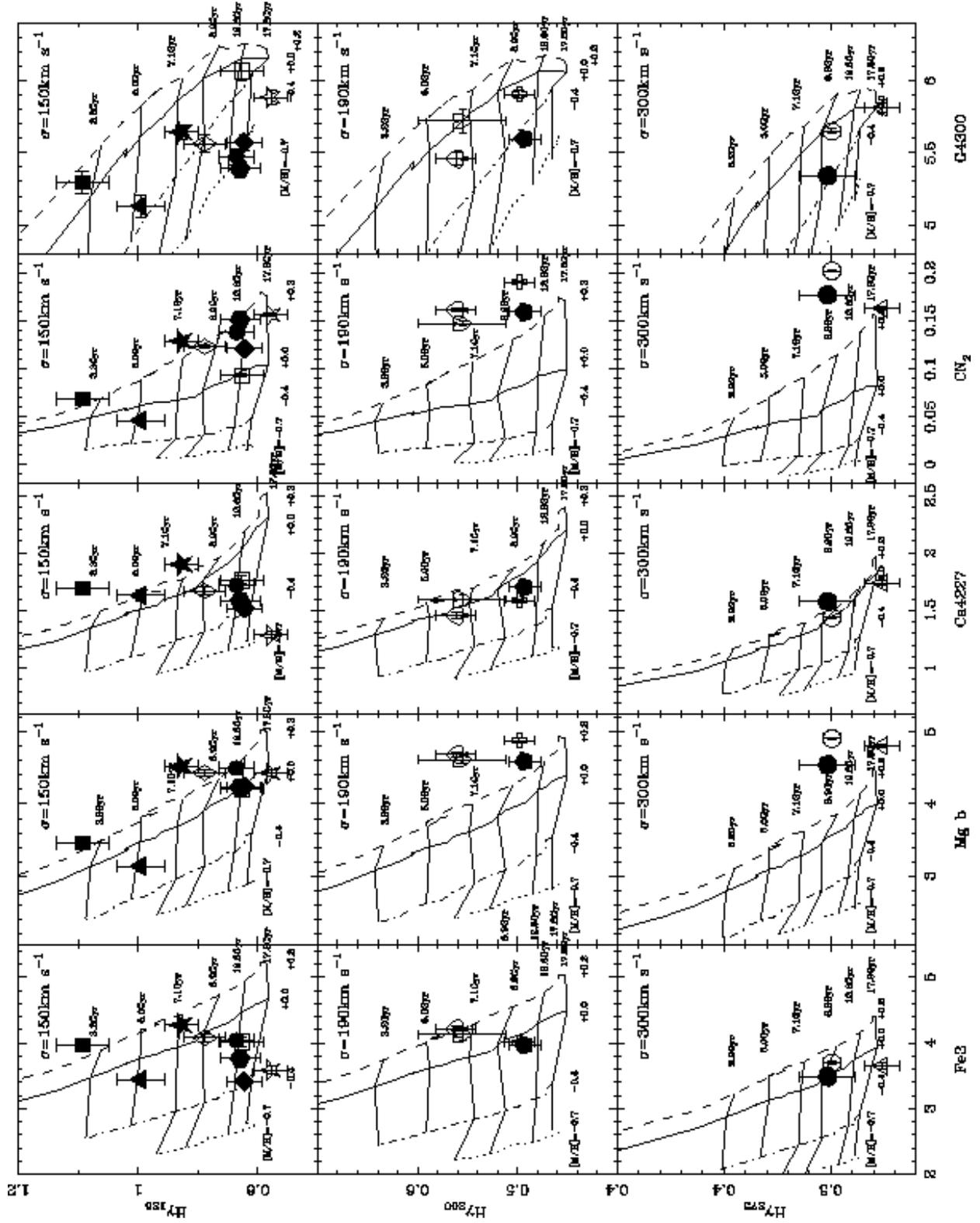}
\caption{
\footnotesize
H$\gamma_\sigma$ vs Fe3, Mg$b$, Ca4227, CN$_2$ and G-band. 
Symbols are the same as in Fig.~\ref{CMR-BLE92}.
\normalsize
\label{Hgamma-Metal-V}}
\end{figure*}

\begin{figure*}[!p]
\epsscale{1.0}
\plotone{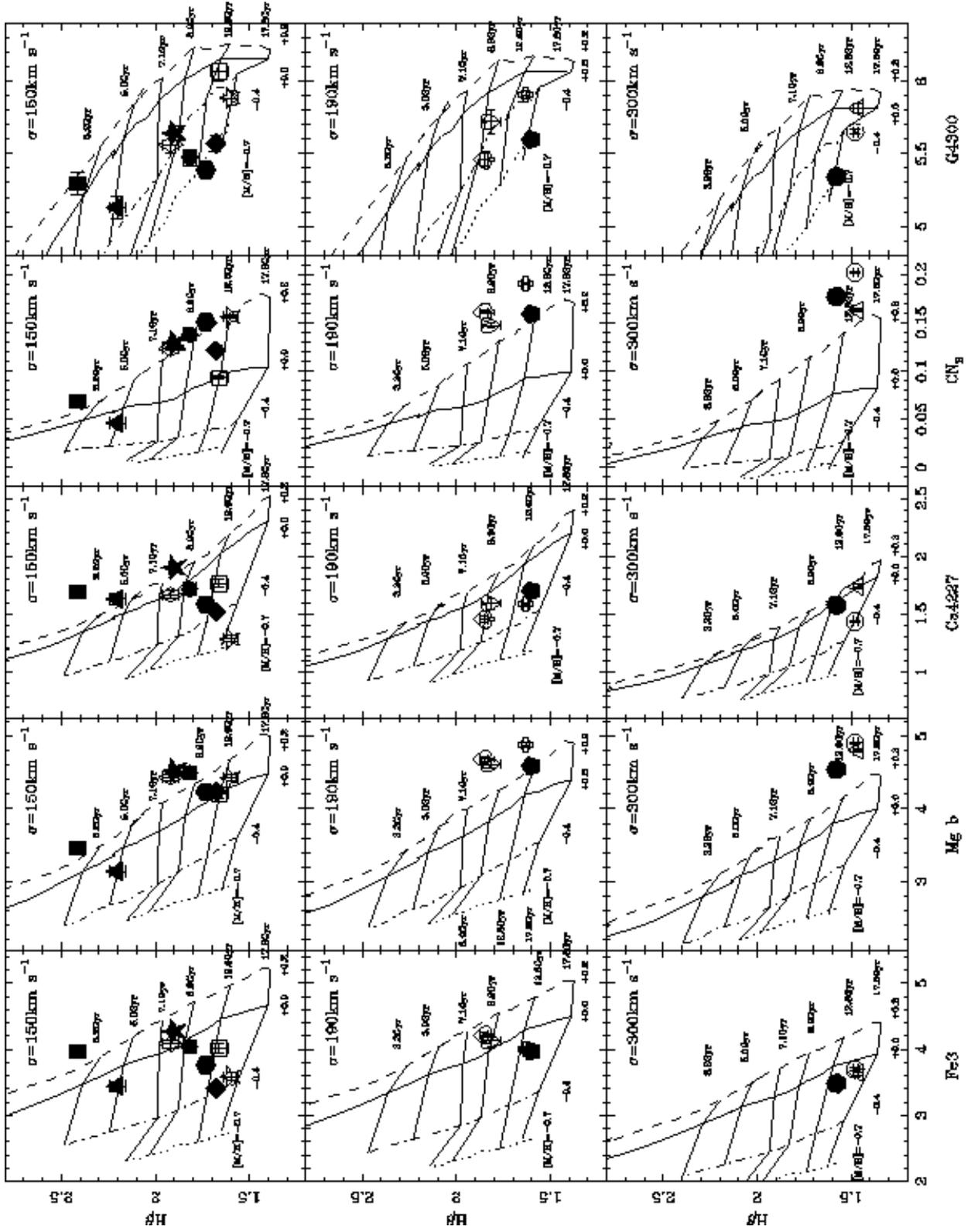}
\caption{
\footnotesize
The Same as Fig.~\ref{Hgamma-Metal-V}, but for H$\beta$. 
Symbols are the same as in Fig.~\ref{CMR-BLE92}.
\normalsize
\label{Hbeta-Metal-V}}
\end{figure*}

\begin{figure*}[!h]
\epsscale{0.4}
\plotone{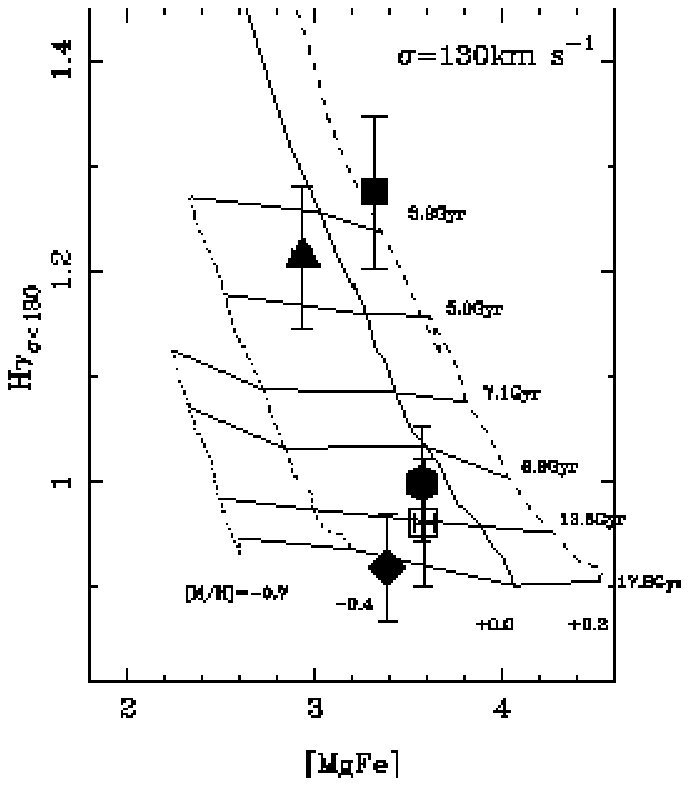}
\caption{
\footnotesize
H$\gamma_{\sigma<130}$ vs. [MgFe] diagram. 
Galaxies with $\sigma <$ 130 km s$^{-1}$ (NGC~4239, NGC~4387, NGC~4458,
NGC~4467 and NGC~4489) are plotted. 
The symbols are the same as in Fig.~\ref{CMR-BLE92}. 
\normalsize
\label{Hgamma130}}
\end{figure*}


\section{Age and Metallicity of Elliptical Galaxies}
Traditionally, a combination of metallic indices and Balmer line indices has been used to
measure the ages of the central regions of observed galaxies (e.g.
Gonz\'alez 1993; Worthey \& Ottaviani 1997). Here we use a number of different
age indicators: H$\gamma_\sigma$, H$\beta$, H$\delta_\mathrm{F}$ as defined by Vazdekis
\& Arimoto (1999), Worthey et al. (1994), and Worthey \& Ottaviani (1997),
respectively. For simplicity, we have divided our sample into three
groups containing galaxy spectra with similar velocity dispersions and
smoothed the model SEDs of Vazdekis (1999) to match the obtained total velocity
dispersion. We chose (1) $\sigma_{\mathrm{measured}}\approx150$ km s$^{-1}$, (2)
$\sigma_{\mathrm{measured}}\approx190$ km s$^{-1}$, and (3)
$\sigma_{\mathrm{measured}}\approx300$ km s$^{-1}$. To allow direct comparisons between
the galaxies having similar velocity dispersions, some small $\sigma$ corrections were
applied : the galaxies with $\sigma_{\mathrm{total}}$
($=(\sigma^2_{\mathrm{galaxy}}+\sigma^2_{instr})^\frac{1}{2}$) $< \sigma_{\mathrm{measured}}$ were convolved
with the appropriate Gaussian to reach a fixed $\sigma_{\mathrm{measured}}$. 
On those spectra we determined our line strength indices. For example,
the spectrum of NGC~4387 with a $\sigma_{\mathrm{total}}$ of 120 km s$^{-1}$ was
convolved with a Gaussian of $\sigma_{\mathrm{extra}}=90$ km s$^{-1}$ to reach a
simulated $\sigma_{\mathrm{measured}}$ 
($=(\sigma^2_{\mathrm{galaxy}}+\sigma^2_{instr}+\sigma^2_{\mathrm{extra}})^\frac{1}{2}$) of 150 km s$^{-1}$.

\subsection{H$\gamma_\sigma$ and Other Age Indicators}
In Fig.~\ref{Hbalmer-MgFe-V}, H$\gamma_\sigma$ (left), H$\beta$ (center) and
H$\delta_\mathrm{F}$ (right) are used as the age indicators, while the often-used
metallicity indicator [MgFe] (Gonz\'alez 1993) is used as the metallicity
index. The velocity dispersion of the galaxies increases in each column from
top to bottom. Note how the model grids change as a function of velocity dispersion. 

The left panel (H$\gamma_\sigma$--[MgFe]) of Fig.~\ref{Hbalmer-MgFe-V} confirms
again that H$\gamma_\sigma$ is well-suited to break the age-metallicity
degeneracy (Vazdekis \& Arimoto 1999). The lines of constant age are almost
horizontal, and the grid of revised Vazdekis (1999) models is more orthogonal 
than the other two index-index diagrams (center and right panels). Moreover, since 
H$\gamma_\sigma$ has much narrower index difinition than the other H$\gamma$ indices such 
as H$\gamma_\mathrm{A}$, metallicity and $\alpha$-enhanced effects should be 
significantly smaller (Thomas, Maraston \& Korn 2004). Therefore, we are able
to derive ages using H$\gamma_\sigma$ under small influence of metallicity, 
or abundance ratios such as [Mg/Fe], though it can be still affected by  
underlying isochrone structure with non-solar abundance ratios. 
Since H$\gamma$ is less
affected by emission lines (see Worthey \& Ottaviani 1997, for example), the
results from H$\gamma_\sigma$ are more reliable in general than those from
H$\beta$. 

Not shown here are diagrams using the other age indicators of Worthey \& 
Ottaviani (1997), i.e., H$\gamma_\mathrm{F}$, H$\gamma_\mathrm{A}$, and H$\delta_\mathrm{A}$, as their 
abilities to disentangle ages and metallicities are similar or even smaller 
than H$\delta_\mathrm{F}$. In Fig.~\ref{Hgamma-Metal-V}, we
show H$\gamma_\sigma$ as a function of several other metal indices: Fe3,
Mg$b$, Ca4227, CN$_2$ and G4300. We also present the H$\beta$ vs. metallicity
indicator diagrams in Fig.~\ref{Hbeta-Metal-V} to demonstrate the disadvantage
of using non-orthogonal diagrams for measuring ages of galaxies with non-solar
abundance ratios. One can clearly see that the age obtained using
H$\beta$--Fe3 is older than that from H$\beta$--Mg$b$, providing H$\beta$ provide less
accurate abundances of various elements. 

For galaxies with $\sigma_{\mathrm{total}}<130$ km/s (NGC~4239, NGC~4387,
NGC~4467 and NGC~4489), the H$\gamma_{\sigma<130}$ index can also be applied
(Fig.~\ref{Hgamma130}). The H$\gamma_{\sigma<130}$ index has a slightly different
wavelength definition; i.e., a wider bandpass for the H$\gamma$ feature than is the case 
for the other H$\gamma_\sigma$ indices. It turns out that the ages derived from
H$\gamma_{\sigma<130}$ for these four galaxies are completely consistent with
those from H$\gamma_{125}$. 

Balmer lines may be filled in with gas emission, becoming weaker and making 
it difficult to measure accurate mean ages. 
Although Gonz\'alez (1993) and Trager et al. (2000a) made corrections to
the H$\beta$ measurements using the [OIII]5007\AA\ emission line, we did not
make any corrections here in this paper. We have investigated whether this correction is
necessary by fitting the whole galaxy spectra with the SSP models to the data. 
If there is a residual at H$\beta$ or at H$\gamma$, it could be attributed 
to the presence of nebular emission. From these residual we found spectra 
that a only few galaxies show emission lines [OIII]4959
and/or [OIII]5007 (NGC~4239, NGC~4489 and NGC~4697, see Fig.~9). However they
are small enough to be ignored for the H$\beta$ and H$\gamma$ age determinations. 

\subsection{Ages from H$\gamma_\sigma$ and H$\beta$}
For each galaxy we determined the ages from H$\gamma_\sigma$ and H$\beta$ by
comparing the data with the SSP models. The resulting ages are given in
Table~\ref{AgeMetal-V}. The errors in the H$\gamma_\sigma$ ages are
dominated mainly by photon noise, while those for H$\beta$ determined
in this way are mostly due to systematic errors coming from the resulting non-orthogonal
index-index model grids. Therefore, while the errors in ages from H$\gamma_\sigma$ can be
improved by longer exposures, this is not straightforward for H$\beta$ ages.
Note that in principle one can obtain better ages from H$\beta$ using more
sophisticated modeling. In Table~\ref{AgeMetal-V} we show the errors which one gets
when using these simple index-index diagrams. We found that in most cases
the youngest ages come from the H$\beta$--Mg$b$ diagram, while the oldest
ages come from H$\beta$--Fe3 diagram. The errors in the H$\beta$ ages listed in
Table~\ref{AgeMetal-V} come from these two diagrams. One can clearly see that
the observed galaxies cover a wide range of ages, from 3 Gyrs to over 15 Gyrs, 
particularly for the galaxies with lower velocity dispersion. 

\begin{figure*}[!h]
\epsscale{0.5}
\plotone{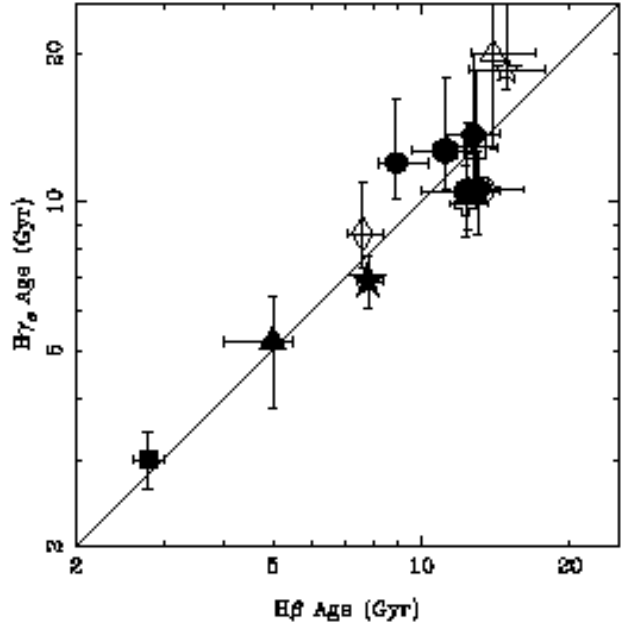}
\caption{
\footnotesize
Comparison between ages from H$\gamma_\sigma$ and H$\beta$. 
Symbols are the same as in Fig.~\ref{CMR-BLE92}. 
\normalsize
\label{AgeAge}}
\end{figure*}

Fig.~\ref{AgeAge} shows the age differences between H$\beta$ and
H$\gamma_\sigma$. They are consistent with each other within the errors, since
H$\beta$ is not strongly affected by emission for our sample. Since [MgFe],
used for deriving the H$\beta$ age, is almost insensitive to the [$\alpha$/Fe]
ratio, we infer that there is no systematic difference between the two ages,
therefore, the H$\beta$ ages are also used in the following discussion. 
We note also that for several velocity dispersions many index -- H$\beta$
diagrams are still somewhat perpendicular, so that also the systematic errors
are not much larger than the errors from photon noise. 


\begin{figure*}[!h]
\epsscale{1.0}
\plotone{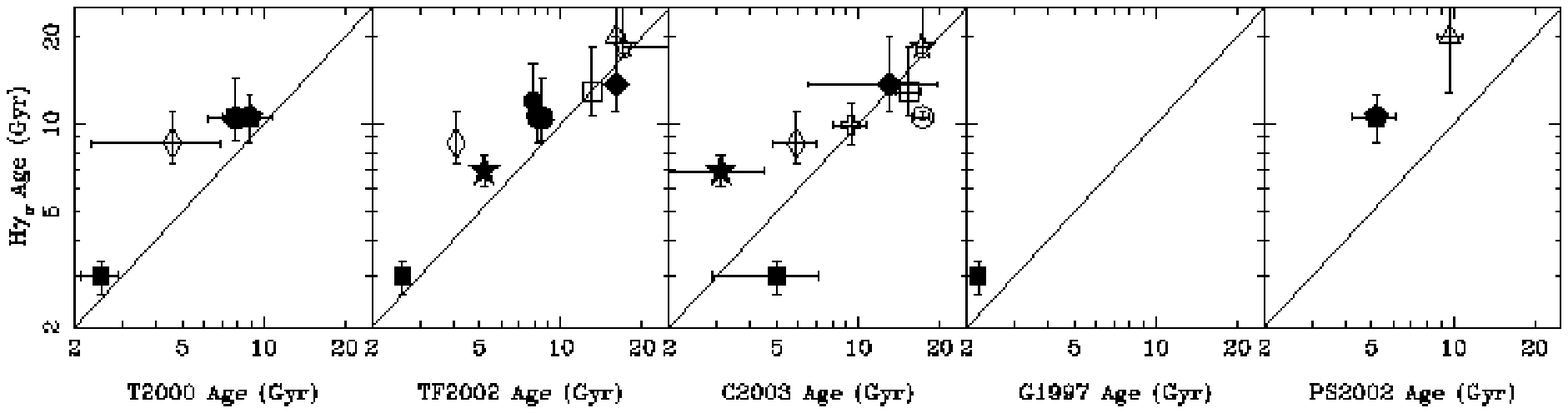}
\caption{
\footnotesize
Comparison between our ages and ages from previous studies. T2000,
TF2002, C2003, G1997 and PS2002 indicate Trager et al. (2000a), Terlevich \&
Forbes (2002), Caldwell et al. (2003), Gorgas et al. (1997) and Proctor
\& Sansom (2002), respectively. Symbols are the same as in Fig.~\ref{CMR-BLE92}. 
\normalsize
\label{AgeAgeLit-V}}
\end{figure*}

Fig.~\ref{AgeAgeLit-V} gives the comparison between our results and the
literature, i.e., Gorgas et al. (1997), Trager et al. (2000a), Terlevich \&
Forbes (2002), Proctor \& Sansom (2002) and Caldwell, Rose \& Concannon
(2003). Gorgas et al. (1997), Trager et al. (2000a) and Proctor \& Sansom
(2002) measured ages from H$\beta$ using Worthey (1994)'s models. In Trager et
al. (2000a), H$\beta$ measurements were corrected for emission using the
[OIII]~5007\AA\ line. Caldwell et al. (2003) used a combination of
Balmer lines (Hn/Fe) and their new SSP models. The ages of Terlevich \&
Forbes (2002) are a compilation of several line index papers (Davies et al.
1993; Carollo et al. 1993; Fisher et al. 1996; Gonz\'alez 1993; Goudfrooij et
al. 1999; Halliday 1998; Kuntschner 1998; Longhetti et al. 1998; Mehlert et al.
1997; Vazdekis \& Arimoto 1999). All age measurements except for those of Vazdekis \&
Arimoto (1999) come from H$\beta$. Our relative ages are consistent with those
of the previous papers, although our results are systematically older. The
systematic difference can be explained in several ways:  1) over correction in
the literature for emission line contamination on H$\beta$.  To correct the
emission effect in H$\beta$ measurements, all the past studies used the 
$\Delta$H$\beta=0.7$[OIII] equation of Gonz\'alez (1993), which had been
derived from the residuals between galaxy spectra and stellar spectra (not SSP
models). We made no correction for emission lines, since our measurements are
based on the H$\gamma$ feature, which is almost free of emission lines due to
residual star formation or planetary nebulae. 2) Worthey (1994)'s model, which
the previous works used, is based on polynomial fitting functions of the
Lick/IDS system. To compare our results with the previous literature, 
it is necessary to convert our measurements of the
indices to the Lick/IDS system. Having done that, we find no systematic 
difference between our measurements and the literature
measurements which use the Lick/IDS system, although the error is large due to 
our relatively small number of reference stars used for the conversion. 
Accordingly, we conclude that the systematic difference in ages does not 
come from differences in the observational configurations used nor from 
data reduction errors. The models of Vazdekis (1999)
employs the Padova Isochrones (Girardi et al. 2000), while the Worthey (1994) model used
the Revised Yale Isochrones (Green et al. 1987). Since the Padova Isochrones
are systematically hotter, this may explain the difference in the estimated ages.
 3) The age estimates corresponding to the H$\beta$
vs. [MgFe] diagram in the Lick/IDS system differ from those estimated using 
flux-calibrated response curves at the specific velocity dispersion of the
galaxy (see Kuntschner et al. 2002). 

We note that NGC~4365 has a very old age, of more than 20 Gyr (but consistent with
14 Gyr within 1$\sigma$). We need S/N/\AA $>$ 350 for galaxies with $\sigma \ge
250$ km s$^{-1}$, while we have achieved in this case S/N/\AA = 250. As a result
the uncertainty in the age is rather large. Note that the H$\beta$ age of NGC~4365
is 14.0 Gyrs, which is consistent with that of Davies et al. (2001). 
Finally note that relative age differences are much more accurate than absolute
ages, due to possible calibration problems in the stellar evolutionary tracks 
(Vazdekis et al. 2001b). 


\renewcommand{\arraystretch}{2.1}
\begin{table}[!h]
\begin{center}
\caption{Age and Metallicity of Virgo Elliptical Galaxies \label{AgeMetal-V}}
\tiny
\begin{tabular}{cccccccccc}
\\
\hline
\hline
NGC  & Age(Gyr)$^a$              & [M/H]$^a$                 & Age(Gyr)$^b$              & [M/H]$^b$ 
     & [Z$_{\mathrm{Fe}}$/H]$^c$ & [Z$_{\mathrm{Mg}}$/H]$^c$ & [Z$_{\mathrm{Ca}}$/H]$^c$ 
     & [Z$_{\mathrm{CN}}$/H]$^c$ & [Z$_{\mathrm{CH}}$/H]$^c$ \\
     &&&&&                         
     & [Z$_{\mathrm{Mg}}$/Z$_{\mathrm{Fe}}$]$^c$             & [Z$_{\mathrm{Ca}}$/Z$_{\mathrm{Fe}}$]$^c$ 
     & [Z$_{\mathrm{CN}}$/Z$_{\mathrm{Fe}}$]$^c$             & [Z$_{\mathrm{CH}}$/Z$_{\mathrm{Fe}}$]$^c$ \\
\hline
4239 &     5.2$^{-1.4}_{+ 1.2}$ & --0.16$^{-0.11}_{+0.10}$ &     5.0$^{-1.0}_{+ 0.5}$ & --0.15$^{-0.05}_{+0.05}$ 
     & --0.14$^{-0.06}_{+0.09}$ & --0.18$^{-0.11}_{+0.12}$ &  +0.20$^{-0.21}_{+0.24}$ & --0.16$^{-0.07}_{+0.08}$ & --0.32$^{-0.16}_{+0.16}$ \\
     &                          &                          &                          &                          
     &                          & --0.04$^{-0.05}_{+0.03}$ &  +0.34$^{-0.36}_{+0.15}$ & --0.02$^{-0.01}_{+0.00}$ & --0.18$^{-0.10}_{+0.07}$ \\
4339 &    12.0$^{-1.9}_{+ 4.2}$ & --0.01$^{-0.09}_{+0.08}$ &     8.9$^{-0.7}_{+ 1.4}$ &  +0.09$^{-0.04}_{+0.04}$ 
     & --0.12$^{-0.06}_{+0.06}$ &  +0.17$^{-0.13}_{+0.09}$ & --0.21$^{-0.11}_{+0.11}$ &  +0.16$^{-0.06}_{+0.06}$ & --0.68$^{-0.12}_{+0.16}$ \\
     &                          &                          &                          &                          
     &                          &  +0.29$^{-0.07}_{+0.03}$ & --0.09$^{-0.05}_{+0.05}$ &  +0.28$^{-0.00}_{+0.00}$ & --0.56$^{-0.06}_{+0.10}$ \\
4365 &    20.0$^{-7.2}_{+10.0}$ &  +0.10$^{-0.09}_{+0.14}$ &    14.0$^{-1.4}_{+ 3.0}$ &  +0.16$^{-0.04}_{+0.05}$ 
     & --0.13$^{-0.07}_{+0.07}$ &  +0.35$^{-0.23}_{+0.20}$ & --0.06$^{-0.09}_{+0.16}$ &  +0.20$^{-0.05}_{+0.09}$ & --0.01$^{-0.39}_{+0.05}$ \\
     &                          &                          &                          &                          
     &                          &  +0.48$^{-0.16}_{+0.13}$ &  +0.07$^{-0.02}_{+0.09}$ &  +0.33$^{-0.00}_{+0.02}$ &  +0.12$^{-0.12}_{+0.00}$ \\
4387 &    12.9$^{-2.2}_{+ 5.5}$ & --0.12$^{-0.09}_{+0.11}$ &    12.9$^{-1.7}_{+ 1.3}$ & --0.12$^{-0.05}_{+0.07}$ 
     & --0.14$^{-0.09}_{+0.08}$ & --0.02$^{-0.11}_{+0.12}$ & --0.20$^{-0.09}_{+0.13}$ & --0.03$^{-0.05}_{+0.07}$ & --0.14$^{-0.34}_{+0.21}$ \\
     &                          &                          &                          &                          
     &                          &  +0.12$^{-0.02}_{+0.04}$ & --0.06$^{-0.00}_{+0.05}$ &  +0.11$^{-0.01}_{+0.04}$ &  +0.00$^{-0.25}_{+0.13}$ \\
4458 &    13.7$^{-2.7}_{+ 6.3}$ & --0.24$^{-0.08}_{+0.08}$ &    12.8$^{-1.6}_{+ 1.7}$ & --0.20$^{-0.03}_{+0.03}$ 
     & --0.37$^{-0.03}_{+0.07}$ & --0.03$^{-0.06}_{+0.13}$ & --0.36$^{-0.06}_{+0.06}$ &  +0.08$^{-0.04}_{+0.05}$ & --0.69$^{-0.10}_{+0.21}$ \\
     &                          &                          &                          &                          
     &                          &  +0.34$^{-0.03}_{+0.06}$ &  +0.01$^{-0.03}_{+0.00}$ &  +0.45$^{-0.02}_{+0.00}$ & --0.32$^{-0.07}_{+0.14}$ \\
4464 &    18.5$^{-1.6}_{+11.5}$ & --0.30$^{-0.06}_{+0.08}$ &    14.9$^{-2.4}_{+ 2.9}$ & --0.21$^{-0.03}_{+0.03}$ 
     & --0.38$^{-0.03}_{+0.13}$ & --0.04$^{-0.08}_{+0.07}$ & --0.66$^{-0.09}_{+0.09}$ &  +0.13$^{-0.04}_{+0.04}$ & --0.66$^{-0.09}_{+0.18}$ \\
     &                          &                          &                          &                          
     &                          &  +0.36$^{-0.11}_{+0.00}$ & --0.26$^{-0.12}_{+0.00}$ &  +0.53$^{-0.07}_{+0.00}$ & --0.26$^{-0.12}_{+0.08}$ \\
4467 &    12.7$^{-2.1}_{+ 5.2}$ & --0.12$^{-0.08}_{+0.10}$ &    11.2$^{-1.6}_{+ 1.5}$ & --0.08$^{-0.05}_{+0.06}$ 
     & --0.23$^{-0.07}_{+0.08}$ & --0.01$^{-0.09}_{+0.15}$ & --0.33$^{-0.05}_{+0.11}$ &  +0.19$^{-0.03}_{+0.06}$ & --0.74$^{-0.14}_{+0.16}$ \\
     &                          &                          &                          &                          
     &                          &  +0.22$^{-0.02}_{+0.07}$ & --0.10$^{-0.00}_{+0.03}$ &  +0.42$^{-0.02}_{+0.04}$ & --0.51$^{-0.07}_{+0.08}$ \\
4472 &    10.5$^{-1.7}_{+ 3.9}$ &  +0.24$^{-0.16}_{+0.17}$ &    12.4$^{-2.4}_{+ 2.1}$ &  +0.11$^{-0.04}_{+0.07}$ 
     & --0.02$^{-0.12}_{+0.13}$ &  +0.52$^{-0.18}_{+0.15}$ &  +0.35$^{-0.41}_{+0.54}$ &  +0.48$^{-0.11}_{+0.12}$ & --0.37$^{-0.30}_{+0.30}$ \\
     &                          &                          &                          &                          
     &                          &  +0.54$^{-0.06}_{+0.02}$ &  +0.37$^{-0.29}_{+0.41}$ &  +0.50$^{-0.01}_{+0.01}$ & --0.35$^{-0.18}_{+0.17}$ \\
4473 &     9.9$^{-1.4}_{+ 1.9}$ &  +0.17$^{-0.09}_{+0.08}$ &    12.3$^{-0.9}_{+ 1.4}$ &  +0.09$^{-0.03}_{+0.05}$ 
     & --0.05$^{-0.04}_{+0.05}$ &  +0.53$^{-0.11}_{+0.09}$ & --0.19$^{-0.08}_{+0.09}$ &  +0.41$^{-0.06}_{+0.06}$ & --0.07$^{-0.21}_{+0.06}$ \\
     &                          &                          &                          &                          
     &                          &  +0.58$^{-0.07}_{+0.04}$ & --0.14$^{-0.04}_{+0.04}$ &  +0.46$^{-0.02}_{+0.01}$ & --0.02$^{-0.17}_{+0.01}$ \\
4478 &     8.6$^{-1.3}_{+ 2.4}$ &  +0.10$^{-0.09}_{+0.12}$ &     7.6$^{-0.5}_{+ 0.8}$ &  +0.16$^{-0.06}_{+0.07}$ 
     & --0.02$^{-0.07}_{+0.08}$ &  +0.27$^{-0.08}_{+0.09}$ & --0.10$^{-0.13}_{+0.17}$ &  +0.18$^{-0.05}_{+0.06}$ & --0.41$^{-0.15}_{+0.17}$ \\
     &                          &                          &                          &                          
     &                          &  +0.29$^{-0.01}_{+0.01}$ & --0.08$^{-0.06}_{+0.09}$ &  +0.20$^{-0.02}_{+0.02}$ & --0.39$^{-0.08}_{+0.09}$ \\
4489 &     3.0$^{-0.4}_{+ 0.4}$ &  +0.26$^{-0.14}_{+0.14}$ &     2.8$^{-0.2}_{+ 0.2}$ &  +0.31$^{-0.08}_{+0.08}$ 
     &  +0.30$^{-0.13}_{+0.14}$ &  +0.28$^{-0.18}_{+0.16}$ &  +0.65$^{-0.21}_{+0.25}$ &  +0.29$^{-0.17}_{+0.16}$ &  +0.11$^{-0.26}_{+0.15}$ \\
     &                          &                          &                          &                          
     &                          & --0.02$^{-0.05}_{+0.02}$ &  +0.35$^{-0.08}_{+0.11}$ & --0.01$^{-0.04}_{+0.02}$ & --0.19$^{-0.13}_{+0.01}$ \\
4551 &     6.9$^{-0.8}_{+ 0.9}$ &  +0.26$^{-0.09}_{+0.08}$ &     7.8$^{-0.4}_{+ 0.6}$ &  +0.21$^{-0.03}_{+0.03}$ 
     &  +0.11$^{-0.06}_{+0.06}$ &  +0.39$^{-0.07}_{+0.08}$ &  +0.38$^{-0.18}_{+0.12}$ &  +0.26$^{-0.04}_{+0.07}$ & --0.20$^{-0.16}_{+0.12}$ \\
     &                          &                          &                          &                          
     &                          &  +0.28$^{-0.01}_{+0.02}$ &  +0.27$^{-0.12}_{+0.06}$ &  +0.15$^{-0.00}_{+0.02}$ & --0.31$^{-0.10}_{+0.06}$ \\
4621 &    10.6$^{-0.3}_{+ 0.4}$ &  +0.35$^{-0.07}_{+0.15}$ &    13.3$^{-0.9}_{+ 2.9}$ &  +0.23$^{-0.02}_{+0.03}$ 
     &  +0.07$^{-0.04}_{+0.05}$ &  +0.68$^{-0.06}_{+0.04}$ & --0.01$^{-0.07}_{+0.13}$ &  +0.53$^{-0.04}_{+0.04}$ & --0.09$^{-0.09}_{+0.09}$ \\
     &                          &                          &                          &                          
     &                          &  +0.61$^{-0.02}_{+0.00}$ & --0.08$^{-0.03}_{+0.08}$ &  +0.46$^{-0.01}_{+0.00}$ & --0.16$^{-0.05}_{+0.04}$ \\
4697 &    10.6$^{-2.0}_{+ 2.0}$ &  +0.11$^{-0.08}_{+0.08}$ &    13.0$^{-1.0}_{+ 1.3}$ &  +0.03$^{-0.01}_{+0.02}$ 
     & --0.08$^{-0.07}_{+0.05}$ &  +0.34$^{-0.12}_{+0.10}$ & --0.11$^{-0.10}_{+0.08}$ &  +0.27$^{-0.05}_{+0.06}$ & --0.45$^{-0.13}_{+0.12}$ \\
     &                          &                          &                          &                          
     &                          &  +0.42$^{-0.05}_{+0.05}$ & --0.03$^{-0.03}_{+0.03}$ &  +0.35$^{-0.00}_{+0.02}$ & --0.37$^{-0.06}_{+0.07}$ \\
\hline
\end{tabular}
\end{center}
\scriptsize
$^{a}$ H$\gamma_\sigma$--[MgFe] diagram. \\
$^{b}$ H$\beta$--[MgFe] diagram. \\
$^{c}$ Metallicities and pseudo-abundance ratios come from the H$\gamma_\sigma$
vs. Fe3, Mg$b$, Ca4227, CN$_2$ and G4300 diagram, respectively. 
\normalsize
\end{table}


\subsection{Metallicity and Abundance Ratios}
By breaking the age-metallicity degeneracy, we can estimate the mean
metallicity [M/H] and a heuristic abundance ratio [Z$_{\mathrm{X}}$/Fe] for the 
elements or molecules X that happen to dominate the absorption of the indices 
plotted in Fig.~\ref{Hgamma-Metal-V}. 

[M/H] is determined from the H$\gamma_\sigma$--[MgFe] and the H$\beta$--[MgFe]
diagram. Thomas et al. (2003) have shown that the
[MgFe] index is a good indicator of the total metallicity [M/H] and almost
independent of [$\alpha$/Fe] (see also Vazdekis et al. 2001b; Bruzual \& Charlot
2003). 

Abundance ratios contain important information about the formation of galaxies, and
are understood to indicate different time-scales for various types of Supernovae
(SNe). The enhancement of $\alpha$ elements by SNe II explosions occurs without
delay from the onset of star formation because their progenitors are massive
stars.  On the other hand, SNe Ia, whose progenitors are white dwarfs of
binaries, mainly supply iron peak elements 1--1.5 Gyrs after the onset of star
formation. As a result, lower [Mg/Fe] indicates that the contribution of SNe Ia is
larger; i.e. star formation is relatively longer (Worthey 1998) or the number
fraction of SNe Ia progenitors is enhanced. Concering CN, both carbon and nitrogen
are ejected from AGB stars whose life time are placed between those of 
SNe II and SNe Ia.
Therefore CN could be another potential indicator for the duration of star
formation (S\'anchez-Bl\'azquez et al. 2003), although the photospheric abundance
of carbon and nitrogen are rather fragile to the convective dredge-up in the
stellar interior and thus may not directly reflect the nucleosynthesis history. 
However, it has been recently found a correlation between the abundance ratio 
of CN with respect to Mg and Fe and X-ray luminosity of galaxy cluster 
(Carretero et al. 2004). 

Since the Worthey (1994) models only have solar abundance ratios, past studies 
(Trager et al. 2000b;
Proctor \& Sansom 2002) used the response functions on the Lick/IDS system of
Tripicco \& Bell (1995) to derive abundance ratio of each element. Since the
method using their response functions is complex and introduces unknown systematic
errors, we did not use it. We therefore 
adopted a more straightforward method using the H$\gamma_\sigma$--metal index
diagrams, i.e., we directly estimated [Z$_{\mathrm{Fe}}$/H],
[Z$_{\mathrm{Mg}}$/H], [Z$_{\mathrm{Ca}}$/H], [Z$_{\mathrm{CN}}$/H] and
[Z$_{\mathrm{CH}}$/H] from the H$\gamma_\sigma$ vs. Fe3, Mg$b$, Ca4227, CN$_2$
and G4300 diagram with the SSP model grids, respectively. For galaxies outside the
model grid, we slightly extrapolated the grid to measure their metallicities. 

Although our method is more straightforward, the estimated abundances
still include uncertainties. For example, it is well known that a line index
has contamination from other elements. Thomas et al. (2003) have
pointed out the [$\alpha$/Fe] bias of stellar libraries using stars in our
Galaxies, i.e., metal poor stars have positive [$\alpha$/Fe]. Another problem is
that the SSP models used here have been made using stellar interior models and
isochrones with solar abundance ratios. One therefore has to keep this in mind,
although it is safe to use them for deriving relative ages as we have shown in this
work.  To differentiate between our heuristically estimated abundances and true
[X/H] abundances, estimated abundances in this paper are written as 
[Z$_{\mathrm{X}}$/H] and so on. 

In Fig.~\ref{Hgamma-Metal-V} we
confirm the well-known result that elliptical galaxies have non-solar abundance
ratios (e.g. Worthey et al. 1992; Trager et al. 2000a). For example, NGC~4365,
NGC~4472, NGC~4473, NGC~4551, NGC~4621 and NGC~4697 show a large enhancement in
[Z$_{\mathrm{Mg}}$/H] and many of our sample galaxies, except for NGC4239, NGC4387
and NGC4489 show significant enhancement in [Z$_{\mathrm{CN}}$/H]. 
[Z$_{\mathrm{Ca}}$/H] is subsolar for most galaxies, which was first pointed out
by Vazdekis et al. (1997), for reasons that are not fully clear yet. One would
expect Ca to behave like other $\alpha$-elements such as Mg, on the basis 
of a nucleosynthetic and chemical evolution sense. Vazdekis et al. (2003)
showed that CaII triplet indices in the infrared are sensitive the global
metallicity for [Fe/H] $< -0.5$ and depend very much on the IMF for 
[Fe/H] $> -0.5$. 


\begin{figure*}[!p]
\epsscale{0.95}
\plotone{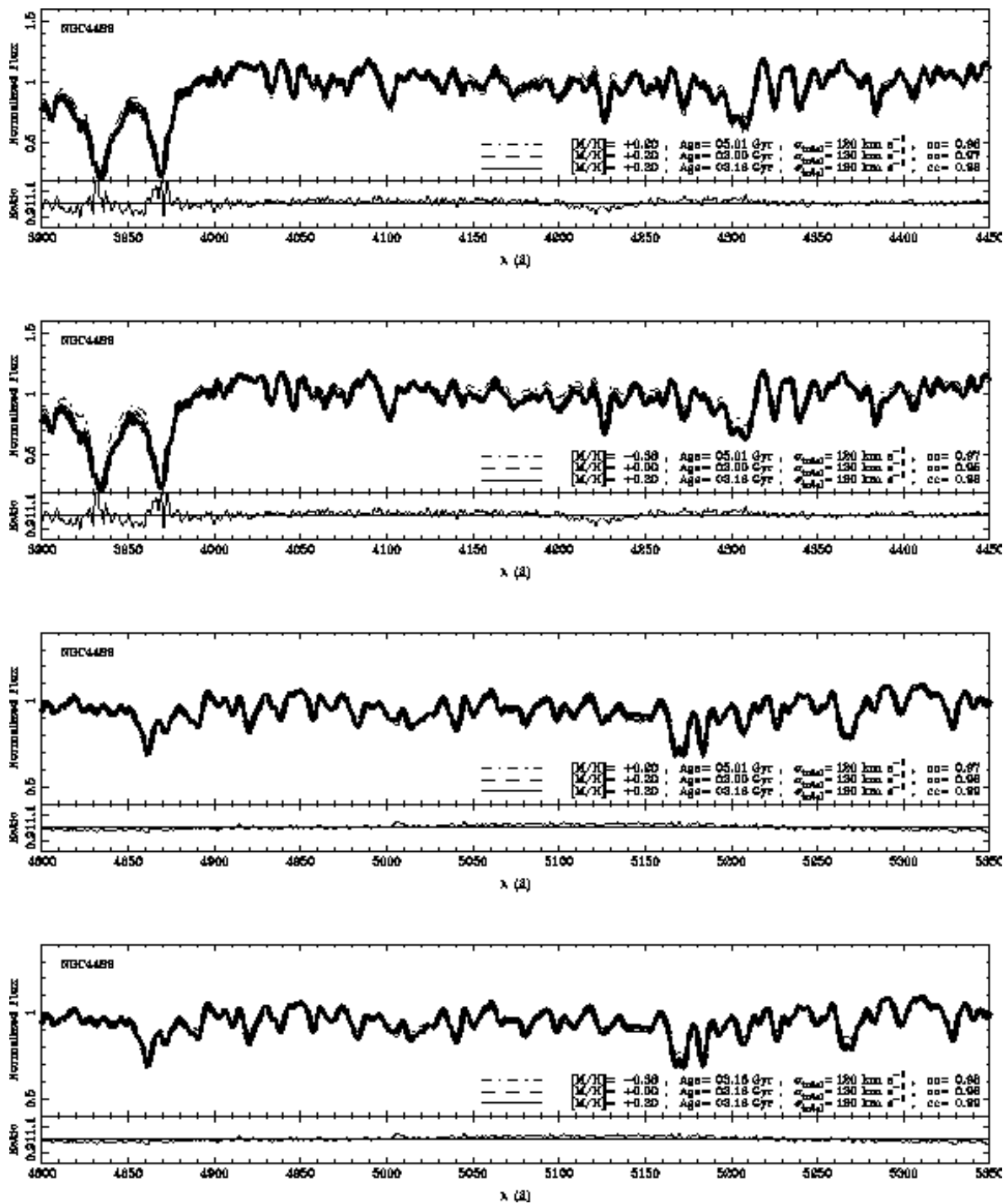}
\caption{
\footnotesize
The full spectral fitting for NGC~4489 using the SSP model spectra of various ages and metallicities. 
Thick and thin lines indicate spectra of galaxies and SSP models, respectively. 
The ratio between the observed spectra and the best fit model is shown in the lower panels. 
Note that we keep the same scale for residuals. 
\normalsize
\label{DetailFit}}
\end{figure*}

\begin{figure*}[!p]
\epsscale{0.95}
\plotone{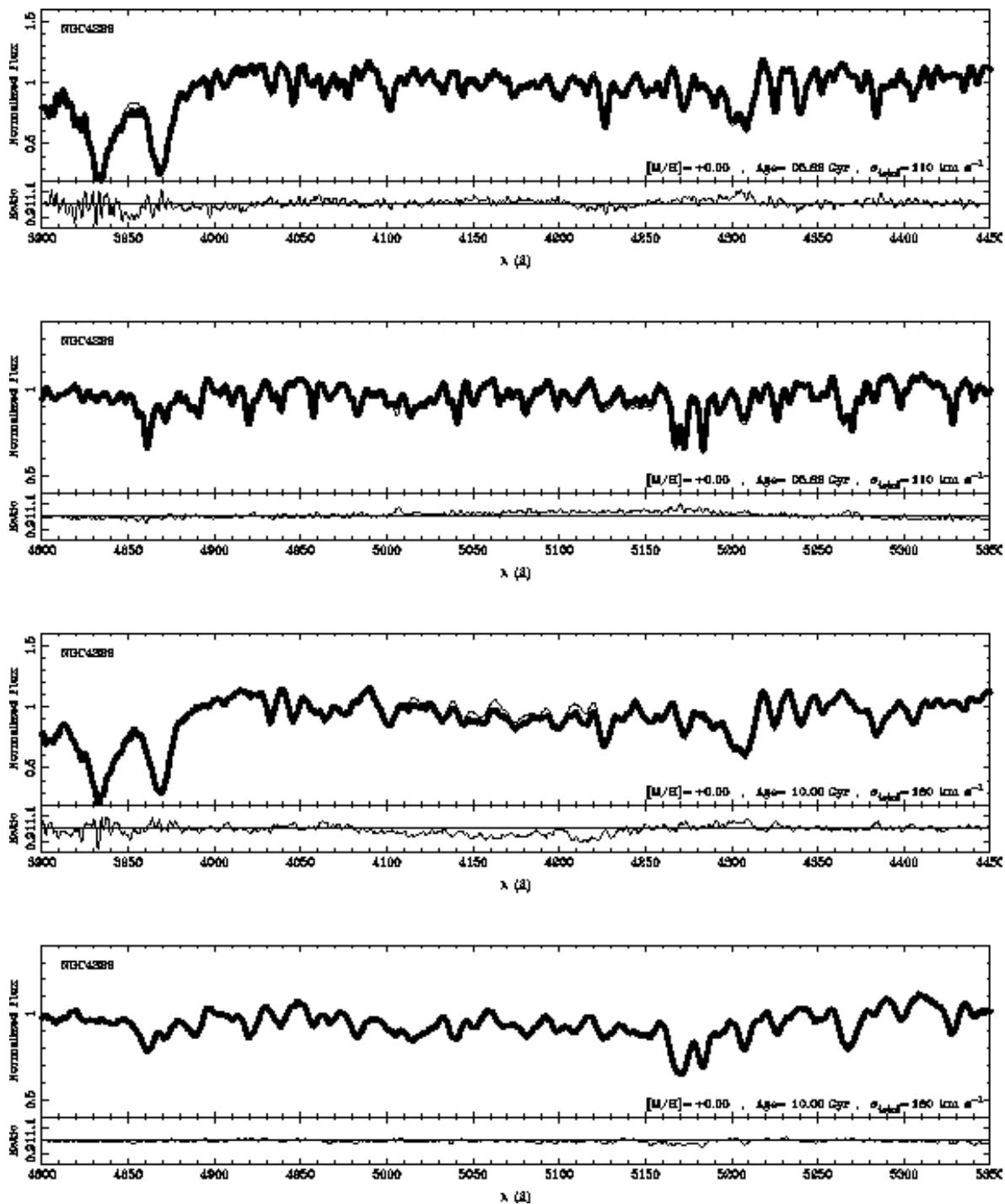}
\caption{
\footnotesize
Spectral fitting for galaxies in the upper panels. The observed spectra 
and the best fit model spectra are indicated by thick lines and thin lines, respectively. 
We present the ratio between the observed spectra and the model spectra 
in the lower panels. We note [M/H], Age and $\sigma_{\mathrm{total}}$ of the model in the
figure. 
\normalsize
\label{Fit1-V}}
\end{figure*}

\begin{figure*}[!p]
\epsscale{0.95}
\end{figure*}

\begin{figure*}[!p]
\epsscale{0.95}
\plotone{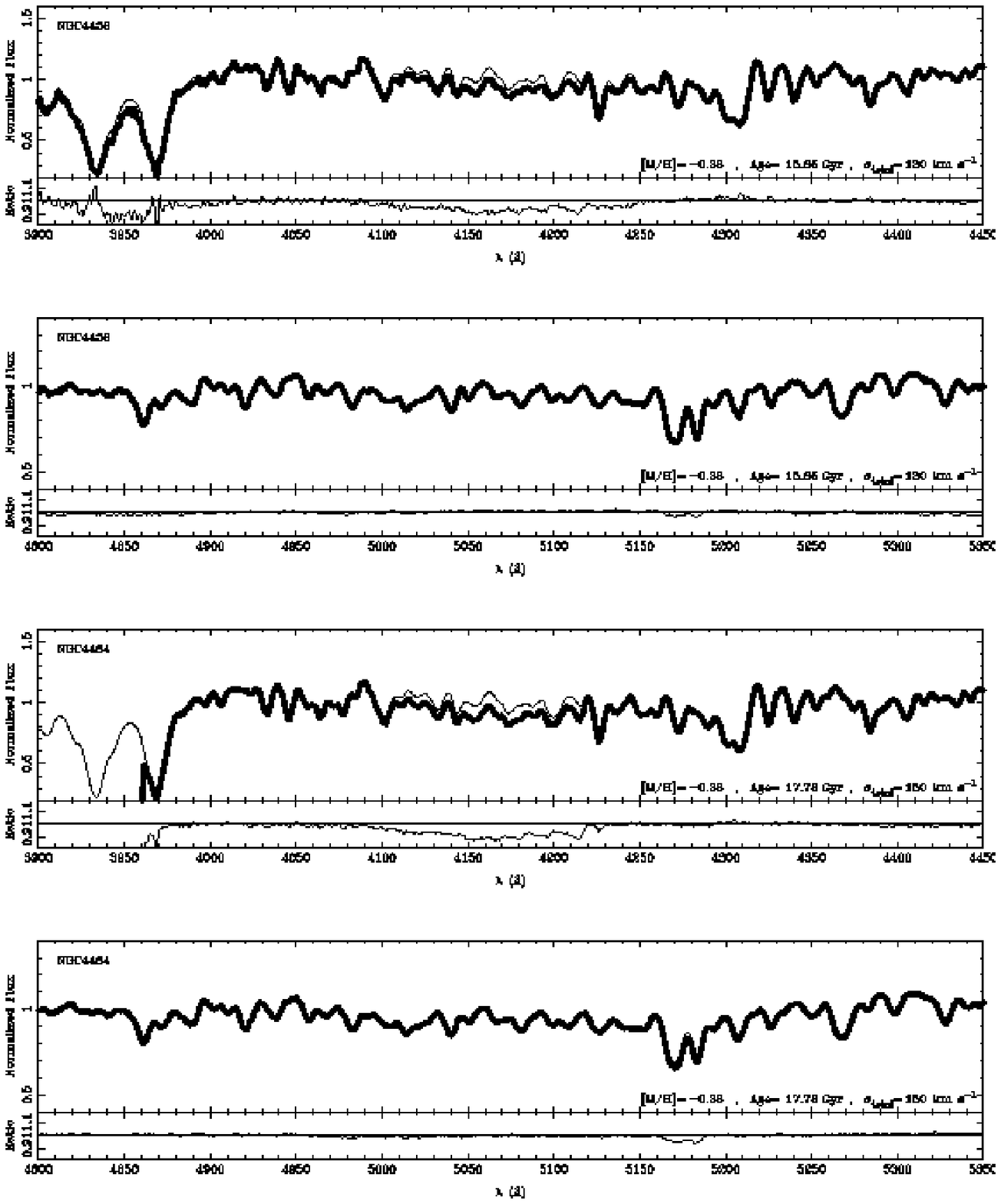}
\end{figure*}

\begin{figure*}[!p]
\epsscale{0.95}
\plotone{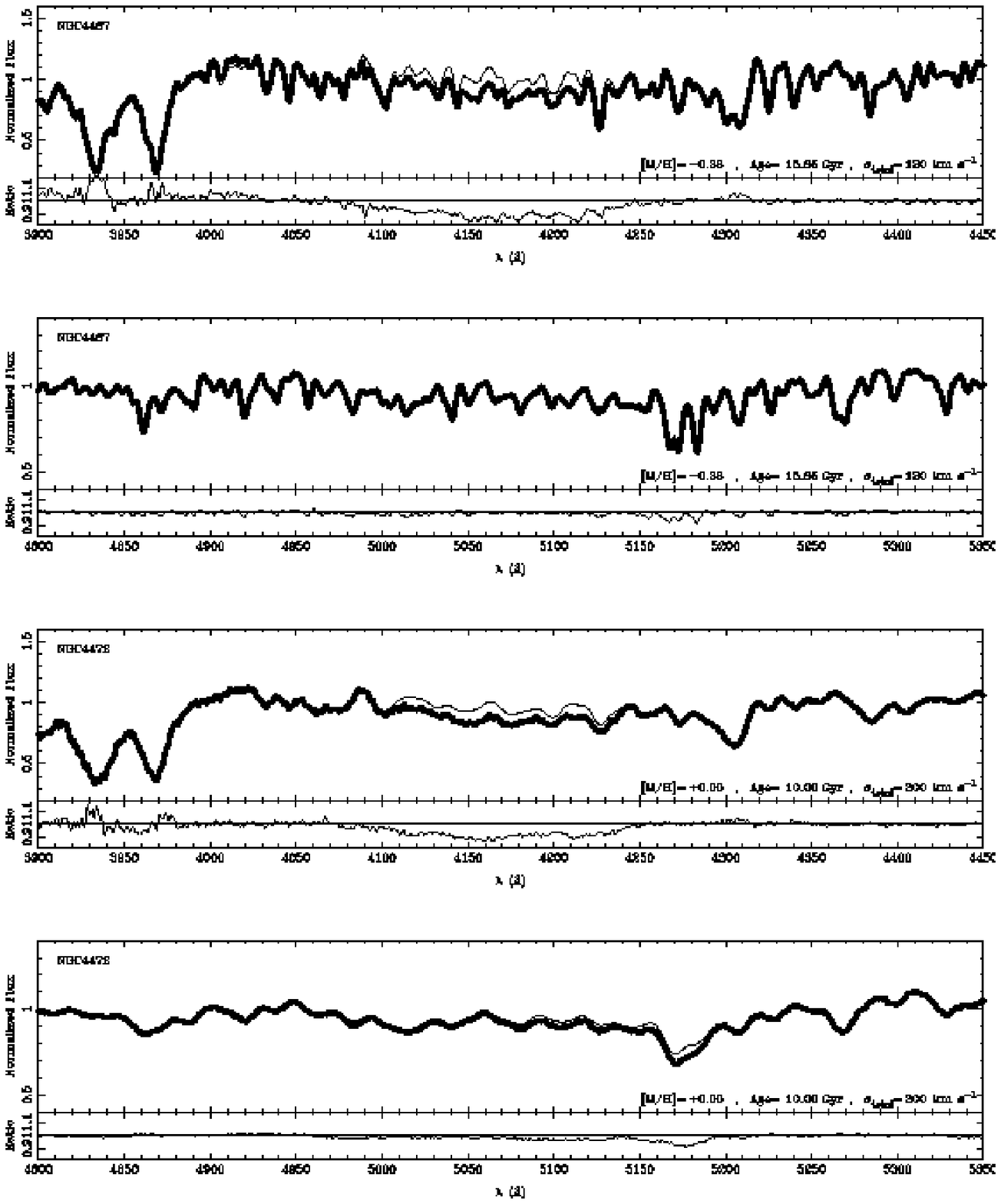}
\end{figure*}

\begin{figure*}[!p]
\epsscale{0.95}
\plotone{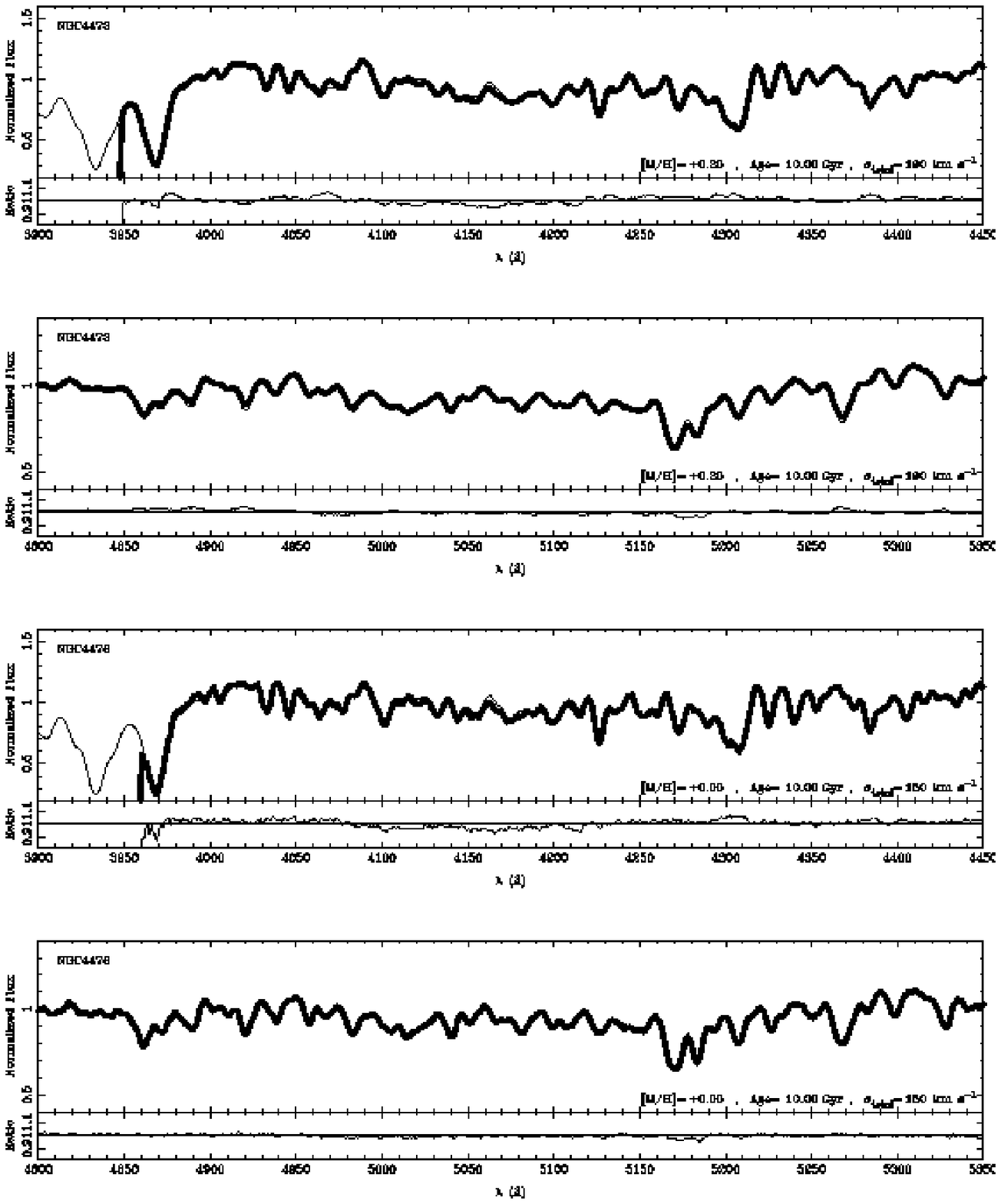}
\end{figure*}

\begin{figure*}[!p]
\epsscale{0.95}
\plotone{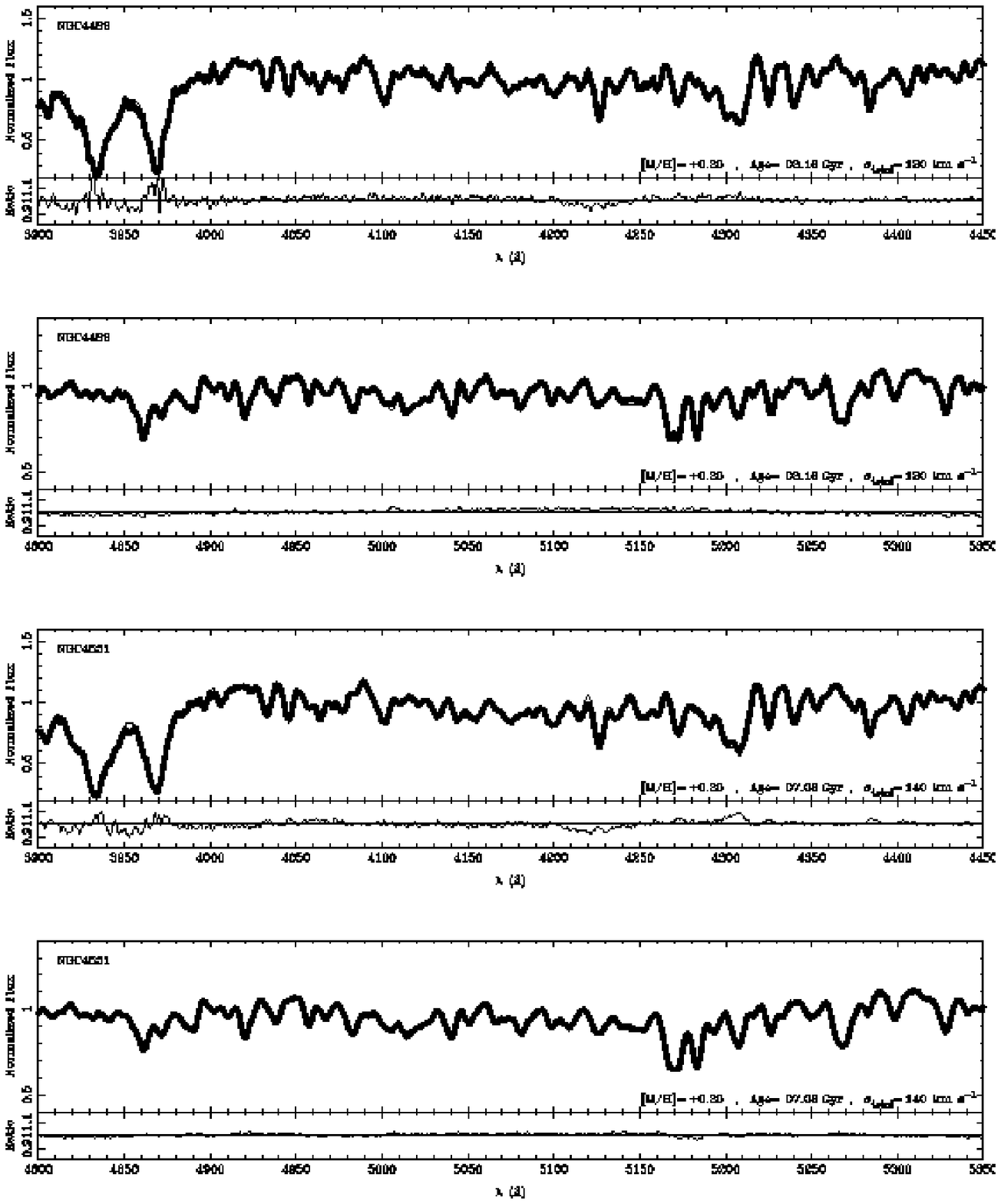}
\end{figure*}

\begin{figure*}[!p]
\epsscale{0.95}
\plotone{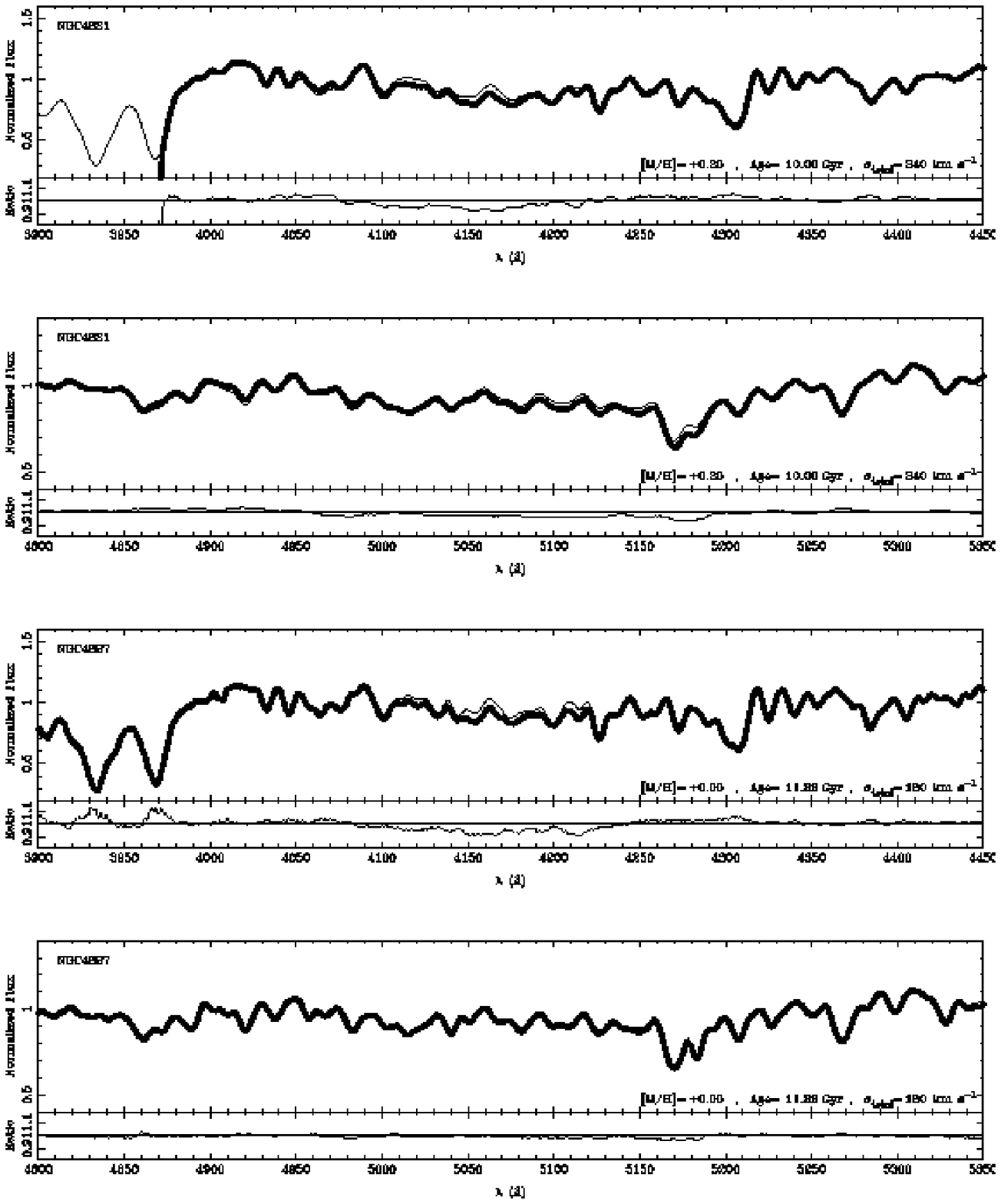}
\end{figure*}


\subsection{Fitting with SSP Models} 
We show the observed and the best matching SSP spectra in Figs.~\ref{DetailFit} \& 9
\footnote{Fig.~9 are available in the electronic version.}. 
Both the observed spectra and the
SSPs are divided by a fitted polynomial of order 2 and normalized at 4350\AA\ and
4950\AA\ for the blue and red spectra, respectively. 
We do not fit wavelength shorter than 4000\AA\ since no data are available there
(WHT) or they have low sensitivity (Subaru).
The observed galaxies are indicated by thick lines, and models by thin
lines. Ages, metallicities and $\sigma_{\mathrm{total}}$ of the models are quoted in each
figure. We also present the ratio of observed spectrum to the SSP model in the lower
panel of each diagram. 

For NGC~4489, we examined the models with several ages and metallicities to
confirm the accuracy of our age and metallicity estimate from the line indices
(Fig.~\ref{DetailFit}). Since the estimated age and metallicity of this galaxy are
3.1 Gyrs and [M/H] = +0.20, respectively, we show the fits with age of 2.00, 3.16,
5.01 Gyrs, and with [M/H] of +0.20, $\pm$0.0, --0.38. For reference, we also show
the cross correlation peak heights derived using the IRAF task ``fxcor'', whose
value of 1.00 means that the perfect fit has been reached. Fig.~\ref{DetailFit} shows
that the model of the same age and metallicity derived by H$\gamma_\sigma$, i.e.,
3.1 Gyrs and [M/H]=+0.20, clearly gives the best fit. 

We have confirmed that most of galaxies show excellent fits to models of ages and
metallicities derived from the H$\gamma_\sigma$--[MgFe] diagram. 
For NGC~4365, whose H$\gamma_\sigma$ age is over 20 Gyrs, 
we estimated that the age is 14--18 Gyrs by fitting of models with various ages. 
One can see the small emission lines of [OIII]4959\AA\ and/or
[OIII]5007\AA\ for NGC~4239, NGC~4489 and NGC~4697, but we cannot see residuals in
either H$\gamma$ or H$\beta$ for any of the galaxies. Hence, we conclude that
emission lines do not affect our age measurements. The overabundance of CN can
also be seen in the fits. The observed spectra of NGC~4365, NGC~4458, NGC~4464,
NGC~4467, NGC~4472, NGC~4621 and NGC~4697 are below the models around the CN
molecular band at $\sim$ 4150\AA. NGC~4464, NGC~4472 and NGC~4621 show stronger
absorption of the Mg triplet line at $\sim$5180\AA. In conclusion, our age and
metallicity measurements agree with the whole spectral fitting and they can be
trusted in our discussion.

\section{Conclusions}
We have obtained very high S/N ($>$100, per \AA) spectra of 8 elliptical galaxies
selected along the CMR in the Virgo cluster with the Subaru Telescope and 6
elliptical galaxies observed with the WHT previously reported in Vazdekis et al.
(2001a), spanning a total luminosity
range of 5.8 magnitude. By using the H$\gamma_\sigma$ method, we have determined
ages, metallicities, and abundance ratios of 14 elliptical galaxies in the
Virgo cluster. 

The H$\gamma_\sigma$ index is almost independent of the galaxy metallicity;
i.e., it can break the age-metallicity degeneracy, therefore the age from the
H$\gamma_\sigma$ index does not depend much on metallicity ([M/H])
or abundance ratios. Although, like all indices, 
H$\gamma_\sigma$ is not completely immune from abundance ratios (Vazdekis et al.
2001b), the relative ages remain secure. The resulting ages show a large spread,
from 3 Gyrs to over 15 Gyrs. We have compared these ages with previous studies
that used the H$\beta$ index and find that the relative ages are consistent. 

We find that Z$_{\mathrm{Ca}}$ (from the Ca4227 index) is following Z$_{\mathrm{Fe}}$,
and that Mg (from the Mg$b$ index) is overabundant with respect to Fe. 
These results are partially known in past studies, however, previous 
approaches were affected by the age-metallicity degeneracy. If one
takes those abundances at face value, it is not easy to reconcile underabundant Ca
with overabundant Mg at the same time, since both Ca and Mg are thought to come
from SNeII. Our results seem to suggest that Ca and Mg are not produced in SNeII
of exactly the same mass range. 

We also fitted the spectra with SSP models finding good agreement with our age and
metallicity determination. Contamination of emission in H$\beta$ and H$\gamma$ is
small enough not to affect our age measurements, while three galaxies (NGC~4239,
NGC~4489 and NGC~4697) have emission line in [OIII]4959 and/or [OIII]5007. Our
fits show the overabundance of some elements in several galaxies: NGC~4365,
NGC~4458, NGC~4464, NGC~4467, NGC~4472, NGC~4621 and NGC~4697 have a strong CN
band at $\sim$4150\AA\ and NGC~4464, NGC~4472 and NGC~4621 show stronger
absorption in the Mg triplet than the fitted SSP model.

In a subsequent paper, based on the ages and metallicities determined here and
together with the velocity dispersions and luminosities, we will discuss the star
formation history of elliptical galaxies along the CMR of the Virgo cluster.

\acknowledgments We thank to the non-anonymous referee, Dr. G.Worthey for detailed 
comments. We thank to Dr.Y.Ohyama, a scientific astronomer of Subaru-FOCAS,
who greatly backed up our observations in detail. We also thank Drs.C.Ikuta and
I.Trujillo who also helped with our observations. Y.Y. is grateful to the Astronomical
Foundation of Japan for traveling support. This work was supported in part by
a Grant-in-Aid for the Scientific Reserch (No.13640230) by the Japanese Ministry
of Education, Culture, Sports and Science. 



\end{document}